\documentclass{article}

\usepackage{arxiv}

\usepackage[utf8]{inputenc} 
\usepackage[T1]{fontenc}    
\usepackage{hyperref}       
\usepackage{url}            
\usepackage{booktabs}       
\usepackage{amsfonts}       
\usepackage{nicefrac}       
\usepackage{microtype}      
\usepackage{lipsum}
\usepackage{graphicx}
\graphicspath{ {./images/} }
\usepackage{xcolor}
\usepackage{dirtytalk}
\usepackage{amsmath}
\usepackage{dsfont}
\usepackage{subcaption}
\usepackage{multirow}
\usepackage{arydshln}

\usepackage{authblk}
\usepackage{lscape}

\usepackage[nodayofweek,level]{datetime}
\newcommand{\mydate}{\formatdate{06}{11}{2023}}

\title{Narratives from GPT-derived Networks of News, \\ and a link to Financial Markets Dislocations}

\author[1, 2, *]{Deborah Miori}
\author[3]{Constantin Petrov}
\affil[1]{Mathematical Institute, University of Oxford, Oxford, UK}
\affil[2]{Oxford-Man Institute of Quantitative Finance, Oxford, UK}
\affil[3]{Fidelity Investments, London, UK}
\affil[*]{\text{Corresponding author: Deborah Miori,} \url{deborah.miori@maths.ox.ac.uk}}
\begin{document}
\maketitle
\begin{abstract}
Starting from a corpus of economic articles from The Wall Street Journal, we present a novel systematic way to analyse news content that evolves over time. We leverage on state-of-the-art natural language processing techniques (i.e. GPT3.5) to extract the most important entities of each article available, and aggregate co-occurrence of entities in a related graph at the weekly level. Network analysis techniques and fuzzy community detection are tested on the proposed set of graphs, and a framework is introduced that allows systematic but interpretable detection of topics and narratives. In parallel, we propose to consider the sentiment around main entities of an article as a more accurate proxy for the overall sentiment of such piece of text, and describe a case-study to motivate this choice. Finally, we design features that characterise the type and structure of news within each week, and map them to moments of financial markets dislocations. The latter are identified as dates with unusually high volatility across asset classes, and we find quantitative evidence that they relate to instances of high entropy in the high-dimensional space of interconnected news. This result further motivates the pursued efforts to provide a novel framework for the systematic analysis of narratives within news.
\end{abstract}



\section{Introduction}

In today's fast-paced digital age, the world is inundated with an unprecedented volume of information, particularly from \textit{news} sources. The sheer magnitude of data generated daily has reached staggering proportions, making it increasingly challenging for individuals to parse and process this information accurately solely through human capabilities. News is constantly flowing from countless channels and platforms, and the need for advanced tools and technologies to sift through this vast sea of data has never been more apparent.
In the realm of financial markets, the potential gain from efficiently handling such an enormous amount of news data, and extract quantitative signals from it, is even more pronounced. Financial markets are highly sensitive to information, and characterising \textit{narratives} within news is surely one task that can enhance our knowledge on news' impact on asset prices, trading strategies, and investor sentiment. 

Some examples of interesting research on the topic follow. In \cite{lit-news1}, the authors investigate the high-frequency interdependent relationships between the stock market and statistics on economic news in the US context. Then, \cite{lit-news2} investigates how news affects the trading behaviour of different categories of investors in a financial market, while \cite{lit-news3} finds evidence that market makers demand higher expected returns prior to earnings announcements, because of increased inventory risks that stem from holding net positions through the release of anticipated earnings news. In \cite{lit-news4}, the authors measure the correlation between the returns of publicly traded companies and news about them as collected from Yahoo Financial News. Then, \cite{lit-news5} tries to quantify how topics discussed within news influence the stock market. The authors of this paper apply a topic modeling technique called Latent Dirichlet Allocation (LDA) \cite{LDA}, in order to extract the keywords of information (i.e. \say{topics}) that synchronise well with trading activity, measured by the daily trading volume. However, LDA is a primitive technique that cannot deliver a systematic and predictive advantage. Relatedly, the very recent research in \cite{topics-finance} investigates the field of topic modeling in the context of finance-related news impact analysis, and further stresses how very limited literature exists. The authors compare three state-of-the-art topic models, namely LDA, Top2Vec \cite{angelov2020top2vec}, and BERTopic \cite{grootendorst2022bertopic}, and show that the latter performs best. The framework focuses on extracting topics and related sentiment of specific stocks, whose time series of returns are in connection investigated. However, the actual identification and interpretation of topics is still of questionable level, and the framework limits itself to facilitate efficient news selection based on topics, but do not properly analyse the latter.

On the other hand, the recent emergence of novel Large Language Models (LLMs), such as OpenAI's Generative Pre-trained Transformer (GPT) ones, clearly marks a significant departure from earlier language processing techniques. These models leverage the power of deep learning and vast amounts of text data to achieve unprecedented levels of language understanding and generation, achieving outstanding improvements in Natural Language Processing (NLP) tasks, and opening countless novel paths of research. The paper in \cite{yu2023temporal} presents a study on harnessing LLMs outstanding knowledge and reasoning abilities for explainable financial time series forecasting of NASDAQ-100 stocks. Then, \cite{gupta2023gptinvestar} leverages GPT parsing of companies' annual reports, to get suggestions on investment targets. On the other hand, \cite{steinert2023linking} shows the potential improvement of the GPT4 LLM in comparison to BERT for modeling same-day daily stock price movements of Apple and Tesla in 2017, based on sentiment analysis of microblogging messages.

There is a clear increasing interest in understanding how to optimally leverage on the capabilities of GPT models, since the focus has otherwise lied on their qualitative responses. For now, quantitative applications have been very constrained in breadth, and we are also not aware of any existing research that tries to leverage GPTs for topic extraction and narrative detection within news. Thus, we believe that our research provides a strongly innovative approach to the solution of such tasks, by leveraging both GPT models and network analysis techniques.

\paragraph{Main contributions.} Our research introduces a novel framework for topic identification within economic news. Indeed, we leverage on GPT3.5 to extract the main entities (both concrete or abstract) of reference for each article in an available corpus, and aggregate entities' co-occurrence among articles in weekly graphs. Studying the resultant set of graphs allows us to identify clusters of entities that can be mapped back to interpretable non-trivial topics. Furthermore, basic metric such as degree and eigenvector centrality of nodes are shown to characterise the evolution over time of central themes of discussion. Beyond that, we propose to consider the sentiment around main entities of an article as a more accurate proxy for the overall sentiment of such piece of text, and describe a case-study to motivate this choice. One final contribution of our study is the investigation of news' features in relation to financial market dislocations. We design attributes that characterise both the local and global structure of news, and their sentiment and interconnections. This allows us to find quantitative evidence of high entropy in the high-dimensional space of interconnected news, when the latter are associated to moments of unusually high volatility across asset classes.

\paragraph{Structure of the paper.} Section \ref{sec:dataN} introduces the data we collect, and some initial related processing. Section \ref{sec:methodologyN} clarifies the theoretical knowledge that is necessary to understand the approach taken, which is mainly comprised by notions from both NLP and graph theory, and on related embedding techniques. Then, Section \ref{sec:resultsN} highlights our analyses on narratives and market dislocations, and describes the results achieved. Finally, we conclude this work with some last remarks in Section \ref{sec:conclusionsN}.


\section{Data}
\label{sec:dataN}

\subsection{Corpus of news}
\label{sec:corpusN}

We download a tractable corpus of news from Factiva\footnote{\url{https://www.dowjones.com/professional/factiva/}} data provider by looking for news written in English, which also belong to the \say{Economics} section of the Wall Street Journal (WSJ).
In this way, we aim at having a set of news that carries a low-noise and focused view on the evolution of news themes that can be of help for financial markets understanding.
\textcolor{black}{We consider approximately four years of daily news, i.e. from January 2020 to October 2023, and aggregate them at weekly level. After pre-processing them to a standard format, we achieve a overall dataset of $197$ weeks with a total of $21,590$ news, with $110 \pm 21$ data points per week (i.e. average number of articles and its standard deviation). Importantly, we considered the week ending on 14th March 2021 as an outlier and dropped it, since we could only download three associated articles from Factiva.}

\subsection{Market dislocations}\label{sec:z-scores-dislocations}

Our aim is to identify and quantify the evolution of narratives within news, but with the further end goal to unravel consequent relationships to the evolution of financial markets.
In particular, we are interested in financial market dislocations, which are often recognised as moments when \say{financial markets, operating under stressful conditions, experience large, widespread asset mispricings} \cite{dislocations}. However, we decide to adopt more a data-driven definition of market dislocations, and consider them as dates when combined shocks to equity, FX, bond, and macro factor risk premium indices occur, i.e. shocks to all the major asset classes.

\begin{figure}[h] 
    \centering
    \includegraphics[width=\textwidth]{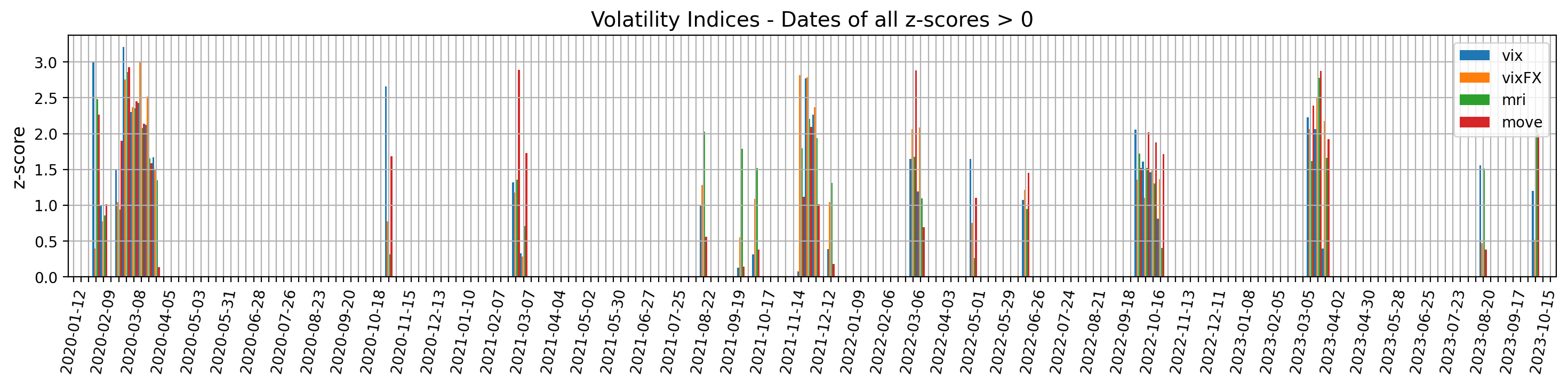}
    \caption{\textcolor{black}{Weeks on which our volatility indices have all positive z-scores, with respect to a rolling window of length $\Delta T = $ 13 weeks. According to the strength of these z-scores, broad market dislocations can be consequently identified.}}
    \label{fig:vol-indices}
\end{figure}

We begin by downloading the following four indices from Bloomberg L.P. at a weekly frequency, with the proposed descriptions taken from its interface:
\begin{enumerate}
    \item VIX Index - \say{The VIX Index is a financial benchmark designed to be an up-to-the-minute market estimate of the expected volatility of the S\&P 500 Index, and is calculated by using the midpoint of real-time S\&P 500 Index option bid/ask quotes.}
    \item JPMVXYEM Index (VIX FX) - \say{J.P. Morgan Emerging Market Currency Implied Volatility Index.}
    \item MRI CITI Index - \say{The Citi Macro Risk Index measures risk aversion based on prices of assets that are typically sensitive to risk. A reading above (below) 0.5 means that risk aversion is above (below) average.}
    \item MOVE Index - \say{The MOVE Index measures U.S. bond market volatility by tracking a basket of Over-the-Counter options on U.S. interest rate swaps.  The Index tracks implied normal yield volatility of a yield curve weighted basket of at-the-money one month options on the 2-year, 5-year, 10-year, and 30-year constant maturity interest rate swaps.}
\end{enumerate}
Then, we compute the related \textcolor{black}{z-scores for a rolling window $\Delta T$ of three months, i.e. $13$ weeks if we assume one year to be made of $52$ weeks}. The z-score is defined as
\begin{equation}
    \text{z-score} = \frac{i - \mu}{\sigma},
\end{equation}
where in our case $i$ is the current value of the index, $\mu$ is its mean over the previous time range $\Delta T$, and $\sigma$ is the related standard deviation. Intuitively, the z-score shows how many standard deviations above the mean the current outcome is.
Figure \ref{fig:vol-indices} proposes the dates for which all our four indices have positive z-score. According to the strength of these z-scores, this can imply that broad dislocations across asset classes were witnessed.

\section{Framework}
\label{sec:methodologyN}


\subsection{Natural Language Processing for Topic Modeling}
\label{sec:methodologyN-GPT}

Natural Language Processing (NLP) has undergone significant advancements in recent years, revolutionising the field of computational linguistics. One of the noticeable but challenging applications of NLP is topic analysis, a technique that involves the extraction of latent themes and subjects from extensive textual corpora. Among the first methodologies proposed for such task, there is Latent Dirichlet Allocation (LDA) \cite{LDA}. 

LDA first transforms each document of a corpus into a set of words with related Term Frequency - Inverse Document Frequency (TF-IDF) value, which is a measure of relative importance of words across documents. This is computed as
\begin{equation}
    \textsc{TF-IDF} = \textsc{Term Frequency (TF)} \times \textsc{Inverse Document Frequency (IDF)},
\label{tfidf}
\end{equation}
where TF is the count of a target word in the current document considered (normalised by the total number of words in the document), and IDF is the inverse of the count of occurrences of such word in the whole set of documents (normalised by the total number of documents). Intuitively, the TF-IDF words' weights per document are lower if the word appears in more documents and so does not carry strong characterisation power, but higher if instead it appears often in just one specific document.
Then, LDA assumes that each document is generated from a collection of topics in some certain proportion, and that each topic is itself a group of dominant keywords with specific probability distribution. 
Consequently, we can try to back-engineer the root topics used to generate a large collection of documents, by analysing the co-occurrence of individual tokens and looking for the joint probability distribution of our given observable and target variables. However, the task of assigning a meaningful label to each set of terms (i.e. forming a topic) must be done by the user. 

Then, a big advancement in topic modeling arose with the introduction of BERTopic \cite{grootendorst2022bertopic}, built upon the Bidirectional Encoder Representations from Transformers (BERT) language model. BERTopic generates document embedding with pre-trained transformer-based language models, clusters these embeddings, and finally, generates topic representations with
a class-based variant of the TF-IDF procedure\footnote{\url{https://maartengr.github.io/BERTopic/api/ctfidf.html}}. However, multiple weaknesses are still characteristic of this technique. The model assumes that each document contains only one single topic, and words in a topic result to be often very similar to one another and redundant for the interpretation of the topic itself. Polysemy (i.e. when words have multiple meanings in different contexts) is also a further challenge. Thus, reliable and systematic downstream applications of BERTopic are considered difficult to achieve.

Generative Pre-trained Transformers (i.e. GPT models) represent a groundbreaking advancement in the realm of NLP and artificial intelligence (AI), due to their remarkable ability to understand and generate human-like text. A GPT model is a decoder-only transformer model of a deep neural network, which uses attention to selectively focus on segments of input text that it predicts to be the most relevant. It is extensively pre-trained and has been proven to achieve groundbreaking accuracy on multiple multimodal tasks \cite{openai2023gpt4}. Despite there have not been related advancements on systematic topic modeling yet, we propose a novel methodology that uses it to achieve indeed such scope.

\paragraph{GPTs and downstream extraction of topics and narratives.} As introduced in Subsection \ref{sec:corpusN}, our news data are descriptive textual content, that we instead aim to analyse quantitatively. Our goal is to extract information on the development of topics and narratives in a systematic but interpretable way. We did experiment with the early topic modeling methodologies described above, but found strong limitations (e.g. labeling the topics from LDA and BERTopic is oftentimes highly subjective and challenging, and such topics can extensively overlap, making it difficult to distinguish between them).
Thus, we decided to leverage on the proven ability of the GPT3.5 model to complete tasks of summary composition, sentiment analysis, and entity extraction \cite{RAY2023121}, to achieve our goal. We could of course have directly questioned GPT models for topics and narratives characteristic of news articles, but their inherent randomness and lengthiness in the formulation of answers makes it difficult to identify comparable and reliable results, to then use for downstream tasks.

GPT3.5 is thus used on our corpus of news, to first do data reformatting.
For each one article, we extract:
\begin{enumerate}
    \item a ranking of the five most important \say{entities} discussed in the text, with related sentiment scores $\in [-1,+1]$,
    \item a ranking of the five most important \say{concepts} discussed in the text, with related sentiment scores $\in [-1,+1]$,
    \item the overall sentiment of the article, both as GPT sentiment score and from the basic VADER technique \cite{vader},
    \item a one sentence summary of the article,
    \item an abstract of the article up to twenty sentences long.
\end{enumerate}
The keywords \say{entities} and \say{concepts} are chosen to either focus more on common and proper nouns, or adding a tuning on abstractions that describe categories of objects, respectively.
Importantly, we employ \textit{prompt engineering techniques} to generate more accurate and useful responses, and lower GPT temperature parameter to $0.2$ to make the outputs more structured. We also iteratively refine our requests, and manually compare a sample of results to the actual news text, to assess the quality of GPT parsing. As an example, we propose here the main prompt used:
\begin{itemize}
    \item \textit{prompt0 = " I am an International Economist. Give me the top five entities mentioned in this text, from the most to least important. Reply as a numbered list using one or two words per entity at most. Next to each entity, provide the sentiment about this entity after a /-symbol strictly as a number between -1 and +1. Do not use words or brackets."}
\end{itemize}
Our hypothesis is that focusing on such keywords of news (that become our fundamental building blocks), and the inherent interconnections that we can define by their membership to one same article, will allow us to identify topics and narratives within news.
Thus, we now introduce the relevant concepts from network analysis to model and analyse such interconnections.


\subsection{Network Analysis}\label{sec:methodologyN-community}

An undirected graph $G$ is a pair $G = (V, E)$, where $V$ is a set whose elements are called vertices or nodes, and $E$ is a set of paired vertices that encapsulate related relationships. The elements $(u,v) \in E$ with $u,v \in V$ are called edges or links, and are characterised by some weight $w_{uv}$. In an unweighted graph, $w_{uv} = 1, \; \forall (u,v) \in E$, while a weighted graph generally has $w_{uv} \in \mathbb{R}^+, \; \forall (u,v) \in E$. If $(u,v) \notin E$, then $w_{uv} = 0$. The \textit{degree} of node $v$ is denoted by $deg(v)$ and computed as
\begin{equation}
    deg(v) = \sum_{u \in V} w_{uv},
\end{equation}
which is often the first measure used to gauge the importance of the different nodes in a graph.

On the other hand, one other well-known measure of centrality (i.e. importance) of nodes is the \textit{eigenvector centrality}.
Relative importance scores are assigned to all nodes in the network, based on the concept that connections to high-scoring nodes contribute more to the score of the node in question. Thus, a high eigenvector score means that a node is connected to many nodes who themselves have high scores.
If we first introduce the adjacency matrix of the graph $G$, which is $\textbf{A} = (w_{uv})$, then the relative centrality score $x_v$ of vertex $v$ can be defined as
\begin{equation}
    x_v = \frac{1}{\lambda} \bigg( \sum_{u \in N(v)} x_u \bigg) = \frac{1}{\lambda} \bigg( \sum_{u \in V} w_{uv} \times x_u \bigg),
    \label{eigv}
\end{equation}
where $\lambda$ is a constant and $N(v)$ is the set of neighbours of node $v$.
Equation \eqref{eigv} can be rewritten in vector form as
\begin{equation}
    \textbf{Ax} = \lambda \textbf{x},
    \label{eigv-matrix}
\end{equation}
for which usually there exist multiple values of $\lambda$ that give a non-zero eigenvector solution.
However, the additional requirement for all entries in the eigenvector to be non-negative implies (by the Perron–Frobenius theorem) that only the eigenvector associated with the greatest eigenvalue is the desired centrality measure. Then, the $v^{th}$ component of the related eigenvector gives indeed the relative centrality score of the vertex $v$ in the network.

Importantly, we highlight that it is common to compute the above and further measures on the \textit{giant component} of a graph. The giant component is the largest connected component of the graph (i.e. the subgraph with highest number of nodes), for which there is a path connecting each pair of nodes belonging to such subgraph.

\paragraph{Community detection - primer.} Another important field of research within graph theory is \textit{community detection}, i.e. clustering tightly connected groups of nodes, which is usually achieved by maximising the modularity $Q$ of the partition proposed.
Modularity $Q$ is computed following \cite{modularity} and measures the strength of division of a network into modules. Mathematically,
\begin{equation}
    Q= \frac{1}{2W} \sum_{uv}{\bigg [}w_{uv}-{\frac {deg(u) \times deg(v)}{2W}}{\bigg ]}\delta (c_{u},c_{v})
\end{equation}
where $W=\sum w_{uv}$, and $\delta (c_{u},c_{v})$ is $1$ if $u, v$ are in the same community (i.e. $c_{u} = c_{v}$), or otherwise $0$. Importantly, modularity allows us to compare partitions of the same network, but it is by no means intended to be compared across different networks.
One of the most popular algorithms for uncovering community structure is the Louvain algorithm \cite{Louvain}. The Louvain method consists of two phases, which are iteratively repeated until a local maximum of modularity is obtained.
Starting from an initialised configuration in which each node is considered as a separate community, then the algorithm iterates through each node and evaluates the potential gain in modularity by moving it to a neighboring community. If the gain is positive, then the node is indeed moved to the community that maximises the gain. The second phase implies instead aggregating the communities identified into super-nodes that generate a new and smaller network, on which we go back to apply the first step of the algorithm, and so on.

\paragraph{Community detection - fuzziness.} Another important branch of community detection algorithms relies on spectral methods. Within such approaches, it is common to consider the connectivity of a graph via its Laplacian matrix $\textbf{L}$, i.e.
\begin{equation}
    \textbf{L} = \textbf{D} - \textbf{A},
\end{equation}
where $\textbf{D}$ is the diagonal matrix summarising degrees of nodes. Since such matrix is positive semi-definite, then it can be decomposed into the product of a real matrix and its transpose. The new matrix can be interpreted as an embedding for nodes in the graph, or further clustered to highlight suggested communities. 
In \cite{fuzzy}, the authors consider the negative Laplacian $\textbf{H}=-\textbf{L}$ of a graph as an encryption of its local structure. In full, 
\begin{equation}
    H_{ij} = 
    \begin{cases}
          w_{ij}, & \text{if}\ i\neq j \text{ and } (i,j) \in E \\
          -d_{i}, & i=j\\
          0, & \text{otherwise}.
    \end{cases}
\end{equation}
Then, they apply to it a diffusion equation evolved by an exponential kernel. This is done in order to extract long-range relationships, and reads
\begin{align}
        \textbf{K} &= \exp{\beta \textbf{H}} = \lim_{y\rightarrow \infty} \Big(1 + \frac{\beta\textbf{H}}{y} \Big)^y, \\
        s.t. \frac{d\textbf{K}}{d\beta} &= \textbf{HK} \text{ with }\textbf{K}(0) = \mathds{1},
\end{align}
where $\beta>0$ controls the degree of diffusion and can be tuned by maximising the modularity score of the optimal partition consequently discovered. The resulting matrix $\textbf{K}$ is symmetric and positive definite, and represents similarities among nodes. It can be then normalised as $K_{ij}^{norm} = \frac{K_{ij}}{\sqrt{K_{ii}K_{jj}}}$, and decomposed via Non-Negative Matrix Factorisation (NMF) as
\begin{equation}
    \textbf{K}^{norm} \approx \textbf{VL},
\end{equation}
where $\textbf{V}^{n\times k}, \textbf{L}^{k\times n} \geq 0$. Matrix $\textbf{V}$ is thus interpreted as a reduced features matrix, whose rows refer to each one of the $n$ nodes in the graph and give associated \say{membership degrees} to a number $k$ of different clusters.
By looking at the strongest membership degree for each node (i.e. highest entry per row), we can then deduce a strict partition of the graph into communities and compute the related modularity score. This is useful in order to tune $\beta$ and $k$ to values that maximise $Q$, and consequently deduce a final optimised clustering.

Importantly, we can also compare the $1^{st}$ and $2^{nd}$ largest probability values $v_{i}^{*}$, $v_{i}^{**}$ of each row, to estimate how stable the label of each node is. This allows us to understand which nodes are best representatives of clusters, or more unstable and overlapping among multiple communities. Specifically, we compute the Stable Index $S_{i}$ for each node $i$ as
\begin{equation}
    S_{i} = \frac{v_{i}^{*}}{v_{i}^{**}},
\end{equation}
which measures indeed the stability of the node in the community assigned. Care is needed in the analysis, since $S_{i}$ can blow up if the denominator approaches $0$.


\subsection{Embeddings}\label{methodology-narratives-embeddings}

As already made explicit, our work relies upon the application of both NLP and network analysis techniques. We extract main entities and concepts characteristic of news thanks to GPT, and these become our fundamental building block for network generation (where the related details will be fully explained in Section \ref{sec:resultsN}). To thoroughly analyse such landscape of data, we will also leverage upon
\begin{enumerate}
    \item the word2vec \cite{mikolov2013efficient} embedding techniques for words,
    \item the node2vec \cite{grover2016node2vec} embedding techniques for nodes in a graph.
\end{enumerate}
The former methodology considers sentences as directed subgraphs with nodes as words, and uses a shallow two-layer neural network (NN) to map each word to a unique vector. The result is that words sharing a common context in the corpus of sentences lie closer to each other.
The latter approach focuses on embedding nodes into low-dimensional vector spaces by first using random walks to construct a network neighbourhood of every node in the network, and then optimising an objective function with network neighbourhoods as input.

\paragraph{Words embeddings - benchmark methodology.} Word2vec \cite{mikolov2013efficient} is a seminal method in NLP for word embedding, which operates on the premise that words with similar meanings share similar contextual usage. This approach employs two core models, i.e. either the Continuous Bag of Words (CBOW) or Skip-gram one. In CBOW, the algorithm predicts a target word from its surrounding context, while Skip-gram predicts context words given a target word. Utilising neural networks, these models generate and adjust word vectors to minimise prediction errors. The resulting high-dimensional vectors encode semantic relationships among words, and they are widely applied in various NLP tasks, such as sentiment analysis, text classification, and machine translation. Despite dating back to 2013, word2vec is indeed still a state-of-the-art methodology for the task of words (and bigrams...) embeddings, since e.g. GPT models are not able to provide such micro-level embeddings.

We can thus identify vectors for our entities extracted from news, by leveraging on pre-trained word2vec models available online, and cluster them. This provides us with a very first benchmark for topic identification and characterisation of their evolution over weeks, before moving to study the effect of graph constructs. 
However, the best pre-trained models available to us tend to be based on old sets of text ($\sim$ up to 2015) and are not frequently updated. Since we do not have the resources to train our own model, this implies that we will surely miss vectors for meaningful words such as \say{Covid-19}, \say{Bitcoin}..., and that the results cannot consequently be considered for more than initial exploratory investigations. In any case, the two pre-trained models we use to generate embeddings are:
\begin{enumerate}
    \item Gensim Google News\footnote{\url{https://github.com/RaRe-Technologies/gensim-data\#models}}, which is a word embedding model based on Google news. A $300$-dimensional vector representation is provided for approximately three million tokens ($\sim$ words).
    \item FinText\footnote{\url{https://www.idsai.manchester.ac.uk/wp-content/uploads/sites/324/2022/06/Eghbal-Rahimikia.pdf}} Skip-gram \cite{fintext}, which is a financial word embedding based on Dow Jones Newswires Text News Feed Database. Again, a $300$-dimensional representation is provided for almost three million tokens.
\end{enumerate}


\paragraph{Nodes embeddings.} Once we introduce an underlying graph structure among entities extracted from news, then a vector representation of the nodes ($\sim$ words) can be generated via the node2vec algorithm \cite{grover2016node2vec}.
Taking inspiration from the word2vec idea of preserving knowledge of the common context windows of a word into its embedding, the node2vec algorithm learns a mapping for the set of nodes $V$ to a low-dimensional feature space by maximising the likelihood of preserving network neighbourhoods of nodes.
Given an undirected monopartite network $G=(V,E)$, node2vec learns a mapping function $f:V \rightarrow \mathbb{R}^d$ that produces a low-dimensional representations of nodes, where $d$ is the number of dimensions of our feature space. Now, let $u \in V$ be a source node and $N_{S}(u) \subset V$ its neighbourhood generated by a sampling strategy $S$. The objective of node2vec is to preserve such network neighbourhoods, which can be formalised as
\begin{equation}
    \max_{f} \sum_{u \in V} \log Pr(N_{S}(u)|f(u)),
\end{equation}
and simplified to 
\begin{equation}
    \max_{f} \sum_{u \in V}\Big[ -\log Z_{u} + \sum_{n\in N_{S}(u)} f(n)f(u) \Big]
    \label{obj}
\end{equation}
with $Z_{u}= \sum_{j\in V} \exp(f(u)f(j))$ as the per-node partition function. The two assumptions needed for this step are 
\begin{enumerate}
    \item conditional independence of seeing neighbourhood nodes given the feature representation, i.e.
    \begin{equation}
        Pr(N_{S}(u)|f(u)) = \prod_{n\in N_{S}(u)} Pr(n|f(u)) ;
    \end{equation}
    \item symmetry of the effect of source and neighbour nodes in feature space, incorporated by choosing
    \begin{equation}
        Pr(n|f(u)) = \frac{\exp(f(n)f(u))}{\sum_{j \in V}\exp(f(j)f(u))}.
    \end{equation}
\end{enumerate}
Then, $Z_{u}$ can be approximated by negative sampling (see \cite{neg-sampling})
and Eq. \ref{obj} is optimised using stochastic gradient descent (SGD).

The above objective allows for interesting flexibility in the definition of the neighbourhood of source node $u$, since this depends on the sampling strategy $S$. Consequently, one can tune the algorithm to either focus on sampling nodes in the same community, i.e. assessing homophily, or nodes with similar roles, i.e. looking at structural equivalence. For the former case, a random walk with Depth-First Sampling strategy (DFS) is initiated from $u$, which samples nodes at increasing distances from the source. In the latter case, nodes are instead chosen via a Breadth-First Sampling (BFS) strategy, which hovers closer to the source node. Of course, intermediates between these two options can also be defined.
Formally, a second order biased random walk $u, v_{1}, ..., v_{l-1}$ of length $l$ and $u,v_{i} \in V$ is generated following the distribution
\begin{equation}
    P(v_{i}=z|v_{i-1}=y) = 
    \begin{cases}
          \frac{\pi_{yz}}{D}, & \text{if}\ (y,z) \in E \\
          0, & \text{otherwise}
    \end{cases}
\end{equation}
where $E$ is the set of edges, $D$ a normalising constant and $\pi_{yz}$ the unnormalised transition probability between nodes $y,z$. Assuming that the walk was on node $x$ before reaching $y$, then $\pi_{yz}$ is given by
\begin{equation}
    \pi_{yz} = 
    \begin{cases}
          w_{yz} \cdot \frac{1}{p_{emb}}, & \text{if}\ \varrho_{xz} = 0 \\
          w_{yz} \cdot 1, & \text{if}\ \varrho_{xz} = 1 \\
          w_{yz} \cdot \frac{1}{q_{emb}}, & \text{if}\ \varrho_{xz} = 2
    \end{cases}
\end{equation}
where $w_{yz}$ is the weight of the edge between $y$ and $z$. The shortest path distance $\varrho_{xz}$ is instead computed between nodes $x$ and $z$. The return parameter $p_{emb}$ defines how likely it is to return to the previously visited node, while the in-out parameter $q_{emb}$ controls how further away from the source node we are inclined to go.
The choice $p_{emb}>max(q_{emb},1)$ ensures that we are less likely to sample an already visited node, while $q_{emb}>1$ biases the walk towards nodes closer to the origin one. Finally, we stress that the number of walks generated from each node should be optimised via hyperparameter tuning.


\subsection{Logistic Regression}\label{narr:logistic-regression}

For completeness, we now briefly recall the logistic regression (also known as logit) statistical model. Logistic regression estimates the probability $P(X)$ of a binary event $Y$ occurring or not, based on a given set $X=(X_1, ..., X_n)$ of $n$ independent variables, i.e. predictors, for the instance under consideration. In our case, we can leverage on such a framework to test which news' features might relate to moment of market dislocations (as per our definition on z-scores).

In logistic regression, probability $0 \leq P(X) \leq 1$ is defined via the logistic function
\begin{equation}
    P(X) = \frac{e^{\beta_0 + \beta_1 X_1 + ... + \beta_n X_n}}{1+ e^{\beta_0 + \beta_1 X_1 + ... + \beta_n X_n}},
\end{equation}
which can be re-written in its logit version as
\begin{equation}
    \log \Big( \frac{P(X)}{1 - P(X)} \Big) = \beta_0 + \beta_1 X_1 + ... + \beta_n X_n.
\end{equation}
In the above, $(\beta_0, ..., \beta_n)$ are unknown regression coefficients that need to be estimated. The method of maximum likelihood is generally chosen for the purpose, in which the intuition is to strongly penalise predictions that lie close to the most uncertain value of $\sim 0.5$. Mathematically, this results in estimating coefficients $(\hat{\beta}_0, ..., \hat{\beta}_n)$ such that the likelihood function
\begin{equation}
    l(\hat{\beta}_0, \hat{\beta}_1) = \prod_{i:y_i=1} P(X_i) \prod_{i':y_{i'}=0} (1-P(X_{i'}))
\end{equation}
is indeed maximised. Then, predictions are made by estimating
\begin{equation}
    \hat{P}(X) = \frac{e^{\hat{\beta}_0 + \hat{\beta}_1 X_1 + ... + \hat{\beta}_n X_n}}{1+ e^{\hat{\beta}_0 + \hat{\beta}_1 X_1 + ... + \hat{\beta}_n X_n}},
\end{equation}
and mapping each $\hat{P}(X)>0.5$ to outcome $1$, while the remaining instances to outcome $0$.
Importantly, the tested predictors should not be correlated with each other to avoid having false relationships suggested by the regression. Then, we desire the estimated coefficients to be of high confidence, i.e. to be significant at least at the $p$-value $<0.05$ level.

Finally, it is often the case that prediction classes are strongly imbalanced, and it is needed to over-sample the minority class in the training set for better results. In our case, we leverage on the well-known SMOTE (Synthetic Minority Over-sampling Technique) algorithm \cite{smote} for the purpose, which generates synthetic perturbations of instances in the minority class to achieve indeed a more general decision region for the prediction of these less frequent events. As a side note, we also mention that other machine learning techniques can be of course used for classification and prediction in a framework such as ours (e.g. classification trees, random forests...). However, logistic regression is indeed the final technique adopted in our work, since we simply aim to complete an initial assessment of whether news structure, especially modelled by our graph construct, can provide any enhancement to our understanding of markets.



\section{Results}\label{sec:resultsN}


\subsection{Word2vec benchmark}

\textcolor{black}{Our data correspond to a set of  $197$ weeks, with $110 \pm 21$ data points per week (i.e. average number of articles and its standard deviation).} For every article in each week, we use basic NLP techniques to systematically identify the five main \say{entities} (and \say{concepts}) with related sentiment, which arise as output of the GPT analysis described in Section \ref{sec:methodologyN-GPT}. This step requires dedicated care, since GPT outputs have some level of instability in their format. In the end, we achieve a structured data set of \textcolor{black}{$107 \pm 21$} data points per week to use for our downstream tasks. Importantly, we also implement a weighting scheme for the sentiment of each keyword $k$, according to its rank of importance within the article. Each sentiment value is thus multiplied by a factor $\frac{1}{rank_{k}}$, where $rank_{k} \in \{1,2,3,4,5 \}$.

We begin by considering both the GoogleNews and FinText pretrained word embeddings, and find the vector representative of each one of our keywords (entities or concepts) identified within news. Unfortunately, only $\sim30\%$ of entities are found to have an associated vector in the GoogleNews model, while this percentage increases to $\sim50\%$ with the FinText model. This is due to a mixture of such models being trained on obsolete data, and inadequate complexity in our keywords being sometimes composed by multiple words. Importantly, \say{concepts} are further found to be strongly higher in complexity and structure rather than \say{entities}, and consequently of lower utility. Thus, our study will focus on \say{entities} as keywords for any following experiment.

We take advantage of the two selected vector embeddings to build a benchmark on the main topics addressed by our news articles. For each embedding model separately and each week of data, we complete the following steps:
\begin{enumerate}
    \item We consider the vectors for the three most important entities (by the GPT ranking) in each article of the week. The full set of five entities is not taken, in order to produce more focused results due to the large percentage of missing vectors.
    \item We cluster the found vectors by performing hierarchical clustering, which iteratively aggregates input coordinates according to some measure of similarity or distance. Since our vectors lie on a $300$-dimensional space, we use the \say{average} method with cosine distance to complete this task (i.e. the distance between two clusters is computed as the average of all cosine distances between pairs of objects belonging to the two clusters).
    \item From the above hierarchical clustering, we form flat clusters so that the original observations in each group have no greater a cophenetic distance than $d_{coph}$. The latter is fixed to $d_{coph}^{GoogleNews}=0.85$ and $d_{coph}^{FinText}=0.80$, for the two models, in order achieve a small and tractable number of clusters (i.e. avoid too granular results).
    \item For each cluster, we consider all the words belonging to it and compute the related centroid by averaging their vectors. We save only clusters with more than one point belonging to them.
    \item To map back centroids to words, we compute the cosine distance between the centroid vector and each one of our word vectors belonging to the cluster. The word that is found to lie closer to the centroid is taken as the representative of the cluster, and by extension as representative of the \say{topic} discussed within the cluster.
\end{enumerate}

\begin{figure}[h]
\centering
\begin{subfigure}{.99\textwidth}
    \centering \includegraphics[width=.99\linewidth]{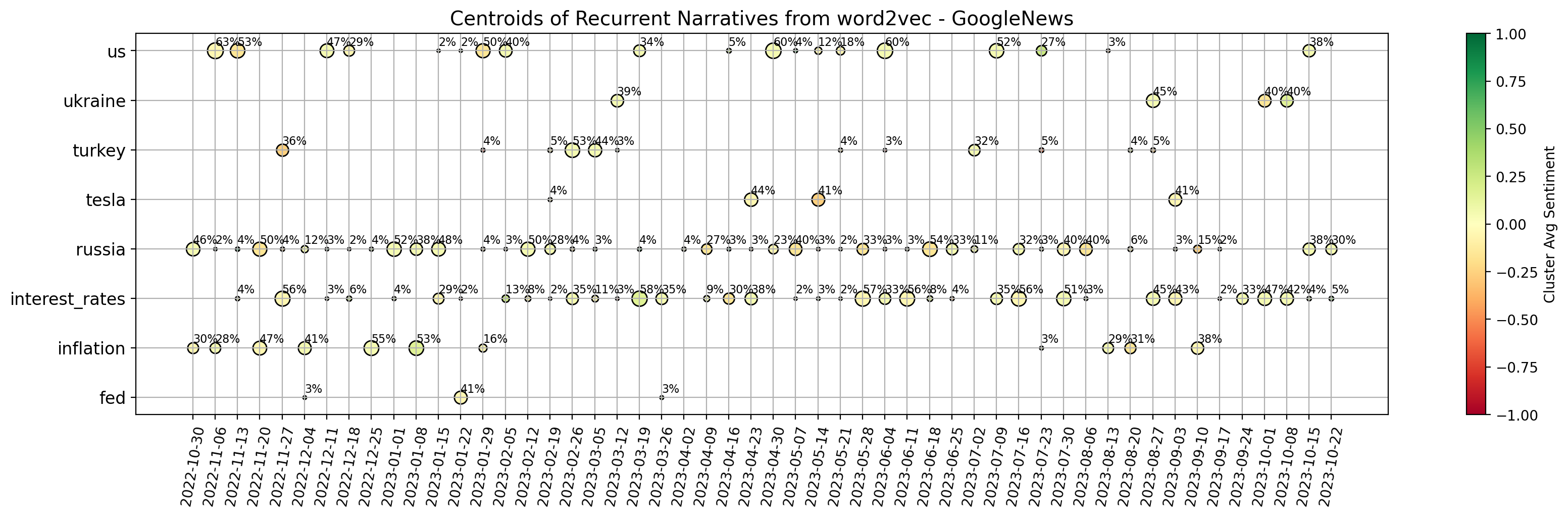}  
    \caption{Results from word2vec GoogleNews pre-trained embeddings.}
    \label{fig:xxx}
\end{subfigure} \vspace{0.4cm} \\
\begin{subfigure}{.99\textwidth}
    \centering \includegraphics[width=.99\linewidth]{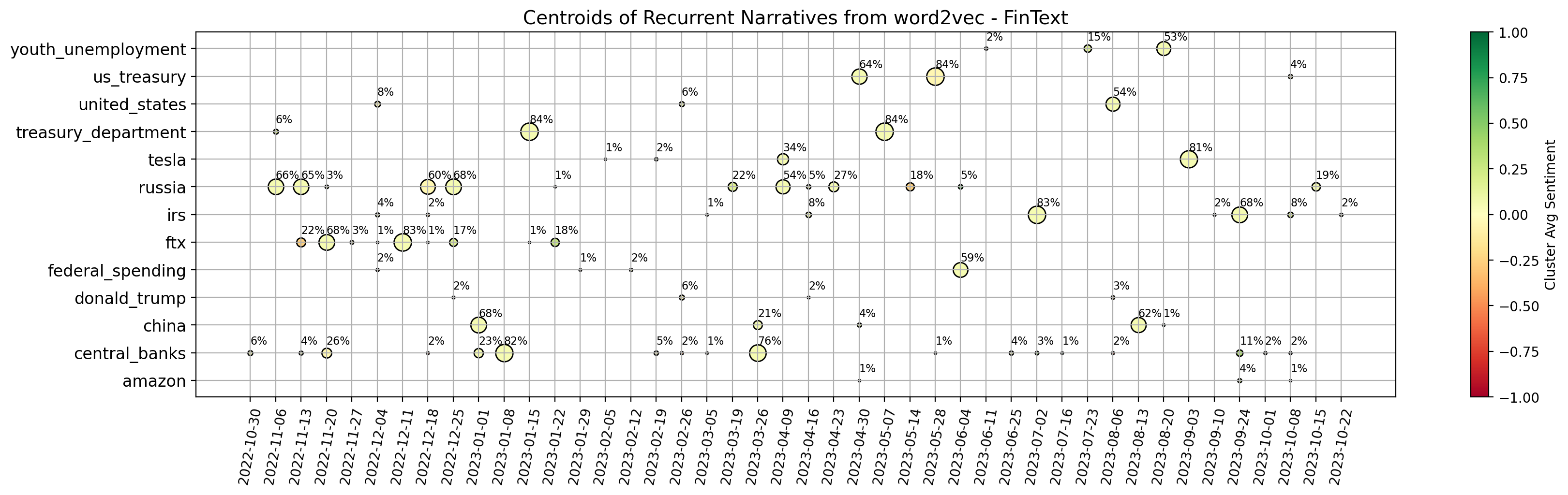}  
    \caption{Results from word2vec FinText pre-trained embeddings.}
    \label{fig:xxx}
\end{subfigure} 
\caption{The proposed plots show the words that are closest to the centroids of clusters of information, which are systemically identified for each week. We plot only centroids that appear more than twice, to produce a clear and more focused representation of the main topics discussed within news. Each point is annotated with the percentage of entities (i.e. words) lying in the related cluster, and coloured by the average sentiment of such entities.}
\label{fig:word2vec}
\end{figure}

\begin{figure}[h!]
    \centering
    \includegraphics[width=0.99\linewidth]{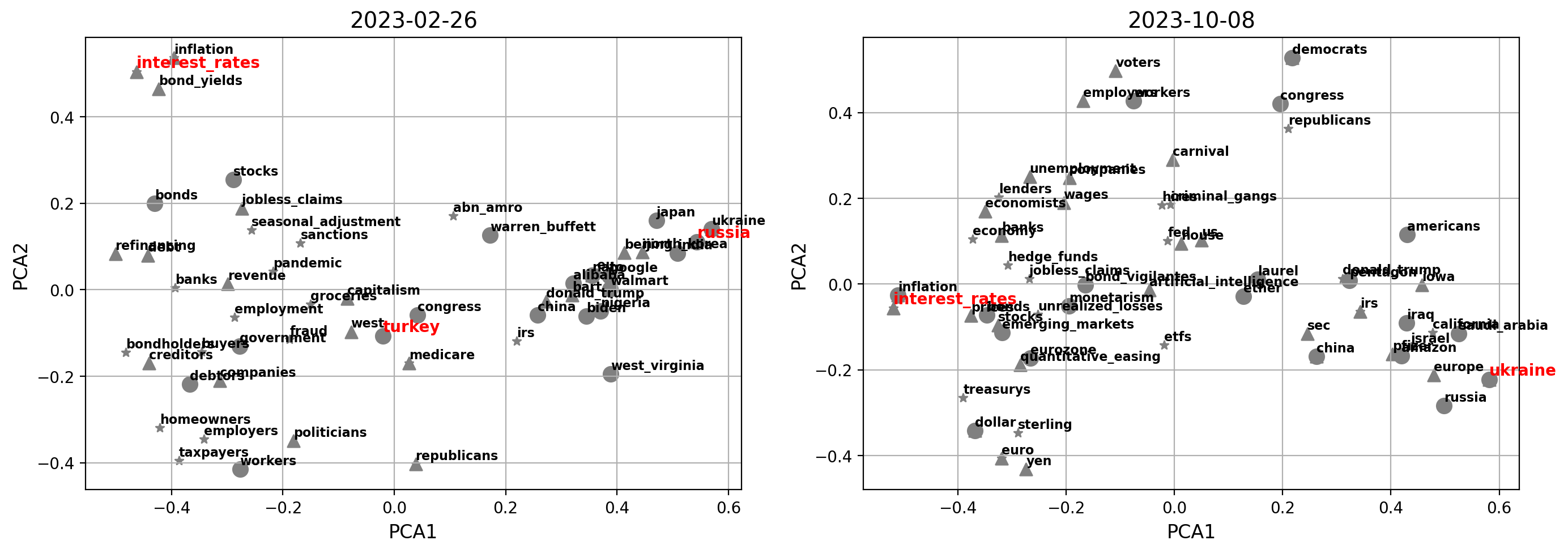}
    \caption{PCA projection of the representative words for two sample weeks, where vectors are recovered from GoogleNews embeddings. Words that are entities of rank 1, 2, or 3, are plotted as circles, triangles, or stars, respectively. We highlight in red the centroids suggested for such weeks from Fig. \ref{fig:word2vec}. Importantly, the variance explained by the two principal components is $\sim 20\%$, while our clustering considers all dimensions of the data.}
    \label{fig:sample-dates-PCA}
\end{figure}

The results for both embedding models are shown in Fig. \ref{fig:word2vec}, for the latest one-year interval defined by weeks ending on 2022-10-30 and 2023-10-22, where we can clearly notice some signs of the different training that the models were subjected to.
The GoogleNews model highlights two main narratives characteristic of the year we are considering, i.e. \say{Russia} (as for the war with Ukraine and associated consequences), and \say{interest rates} (due to continuous hiking of central banks). Interestingly, \say{inflation} is also a topic of main concern especially at the beginning of the data sample, and it is accurate to see that \say{Turkey} is signalled in February 2023 (when indeed a disastrous earthquake unfortunately happened). These results are in line with the fact that the model was trained on a corpus of news. On the other hand, the FinText model provides more disperse results. FinText is focused on financial language and related companies' data, and indeed we see that e.g. the FTX collapse of November 2022 is well identified. \say{Russia} and \say{central banks} are also hinted as recurrent and meaningful centroids of information, but the instability and noise within outcomes is significant. For completeness, Fig. \ref{fig:sample-dates-PCA} shows the Principal Component Analysis (PCA) projection of points for sample dates 2023-02-26 and 2023-10-08, where the vectors come from the GoogleNews embeddings. The first two PCA components explain together $\sim 20\%$ of the variance of the data, implying that the proposed representation must be interpreted with care. If we compare the centroids highlighted in Fig. \ref{fig:word2vec} with the current plot, we anyways see some reasonable (despite noisy) structure identified. However, it is clear that the results can be unstable and have difficulties in unravelling deeper shades of information.

Overall, we conclude that the available pre-trained word embeddings allow us to generate an interesting (despite very basic) initial benchmark on expected central words for topics identification in the given time range. But as highlighted, many problems come with this methodology. We will soon proceed to introducing our novel graph-based methodology for narratives identification, which further allows us to assess whether a topic exists on its own, or is interconnected with other topics. However, we first briefly provide some meaningful remarks on the choices taken to account for the sentiment within our news.

\vspace{1cm}

\subsection{Sentiment choice}

When analysing news, it is of clear importance to be able to accurately assess their related sentiment, especially if one desires to connect them to the state of financial markets. We believe that simply considering the overall sentiment of an article is restrictive, and decide to leverage on what GPT extracts as the sentiment surrounding the articles' main entities.
Thus, we provide now a brief motivating example that supports our approach. 

\begin{figure}[t]
\centering
\begin{subfigure}{.33\textwidth}
    \centering \includegraphics[width=.99\linewidth]{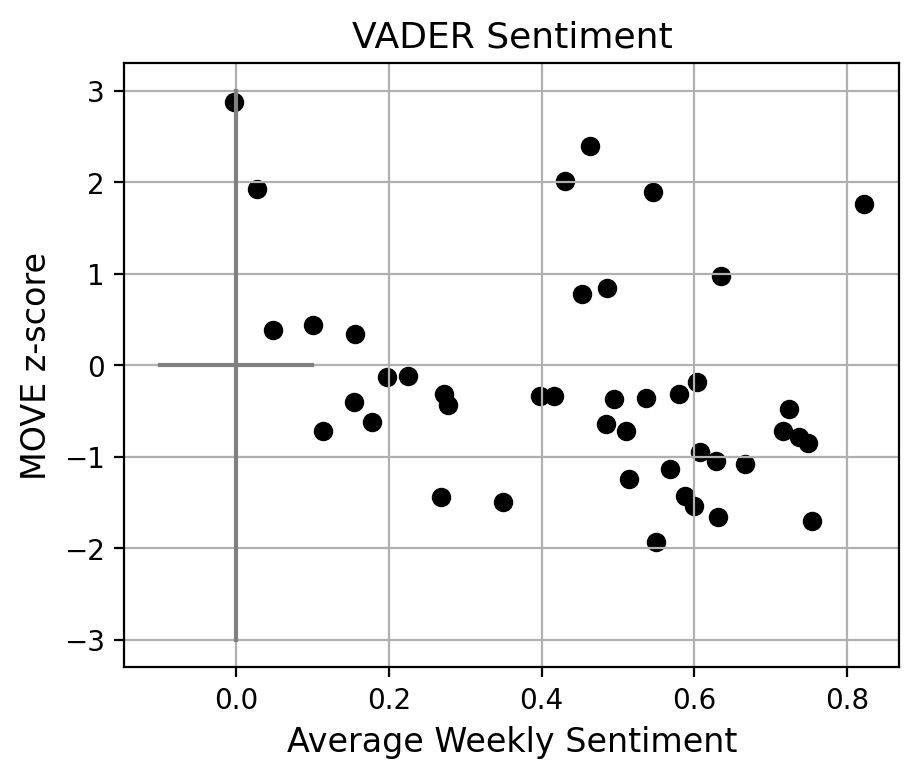}  
    \caption{A positive bias is detected.}
    \label{fig:xxx}
\end{subfigure} \hfill
\begin{subfigure}{.33\textwidth}
    \centering
    \includegraphics[width=.99\linewidth]{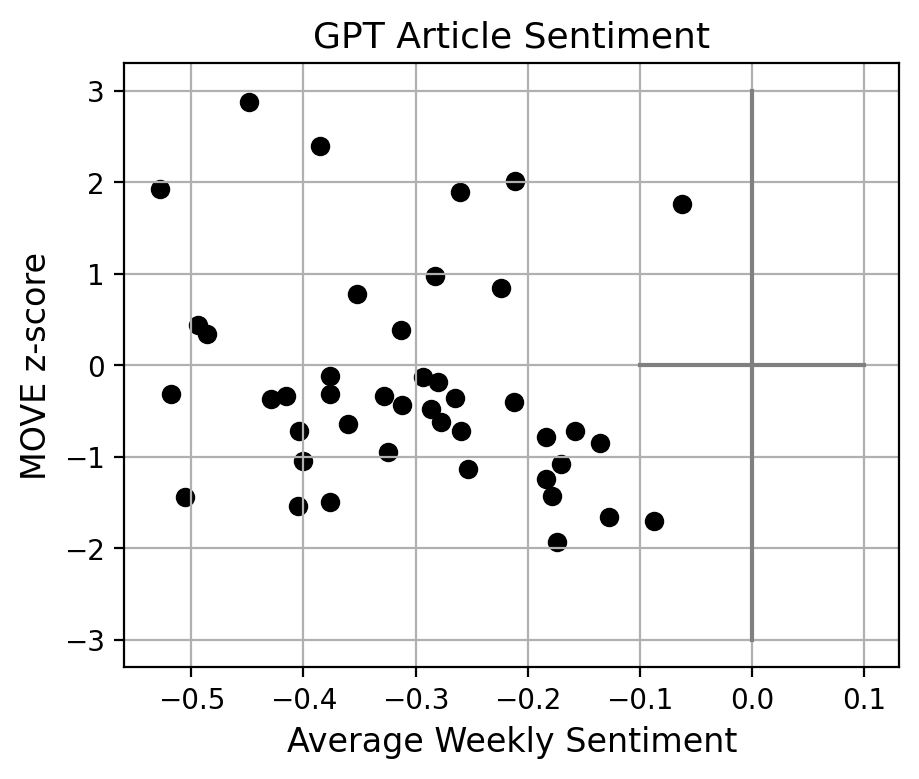}  
    \caption{A negative bias is detected.}
    \label{fig:xxx}
\end{subfigure} \hfill
\begin{subfigure}{.33\textwidth}
    \centering \includegraphics[width=.99\linewidth]{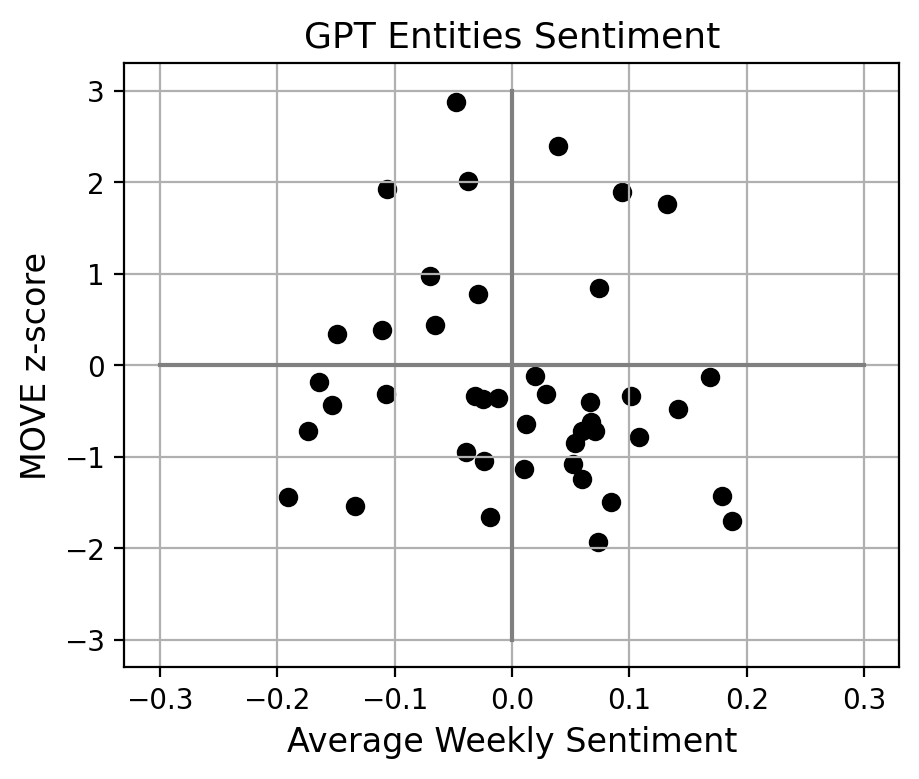}  
    \caption{Improved diagonal relationship. }
    \label{fig:xxx}
\end{subfigure}
\caption{Weekly MOVE Index z-scores against average sentiment of news on each week, for subsets of articles relevant to the bond market. The primitive VADER sentiment analysis technique shows a positive bias, while GPT sentiment calculated for whole articles averages to an opposite negative bias. Since the period considered is not uniformly negative, we are instead satisfied with our proposed entities-based sentiment. This is indeed much more symmetric, despite carries a tail of noise in the third quadrant that we believe spills from the overall negative GPT bias.}
\label{fig:figSentiment}
\end{figure}

\textcolor{black}{Our latest one year of data, i.e from the week ending on 2022-10-30 to 2023-10-22, concerns a strongly focused period of high (but oscillating) concern on themes such as interest rates and inflation. Thus, we use such data in the current experiment.} We subset related articles to the ones where the first or second most important entities are in the set $\{$\say{interest rates}, \say{inflation}, \say{Federal Reserve}, \say{European Central Bank}, \say{Bank of England}$\}$. Then, we consider our weekly z-scores computed for only the MOVE Index, which indeed focuses on the bond market. For each week with more than $15$ related articles, we average the sentiment (always weighted by rank) of the main entities extracted for the article, and then average over articles. Due to the weighting applied, the maximum mean achievable is $\frac{1+0.8+0.6+0.4+0.2}{5}=0.6$, and thus we multiply each result by a factor $0.6^{-1}$. Similarly, we also average over Vader and overall GPT sentiment on the articles for each week, and visualise the results in connection to MOVE z-scores. Figure \ref{fig:figSentiment} shows the trends discovered. VADER is one of the most primitive sentiment analysis techniques, and indeed it reveals a strong positive bias. On the other hand, it is interesting to see that GPT tends to constantly assign an overall negative sentiment to the articles related to our theme under investigations, despite this is not representative of the oscillating level of concern in the actual markets and economy. Thus, we are satisfied to see that our weighted GPT sentiment calculated from entities occupies mainly the second and fourth quadrant of the plot, as desired. There is still a tail of negative sentiment with improving bond market (i.e. negative MOVE z-score), but we explain this as a spillover from the overall negative GPT bias.

\subsection{Graph construction and initial measures}

We can now proceed to introducing our novel graph-based methodology for narratives identification, which further allows us to assess whether a topic exists on its own, or is interconnected with other topics. As already hinted, the building blocks of our methodology are the ranked entities extracted by GPT from each article. For each week of news available, we thus generate a related representative graph $G=(V,E)$ as follows:

\begin{enumerate}
    \item The set of nodes $V$ is constructed from the union of all entities extracted from the articles $a \in A$ of the week. An attribute is also assigned to each node, which is the average of the (rank-weighted) sentiment extracted by GPT around such entity across articles.
    \item For each pair of entities $u,v \in V$ that belong to the same article $a$, an edge $(u,v) \in E$ is created. Such edge has a weight $w_{a, uv}$ given by the product of the inverse $rank$ of the two entities in the article, i.e.
    \begin{equation}
        w_{a, uv} = \frac{1}{rank_{a,u}} \times \frac{1}{rank_{a,v}}.
    \end{equation}
    This choice is taken to better account for the strength of connection among words, implied from their importance and focus within the article. 
    \item Multiple edges across the same two nodes are aggregated, and the associated weights summed to give the final weight $w_{uv} = \sum_a w_{a, uv}$ of the link among each pair of nodes $u$ and $v$. A threshold is then applied, and all edges with $w_{uv} \leq \frac{1}{2} \times \frac{1}{3}$ dropped. Explicitly, we are requiring that the sum of weights among each pair of nodes is strictly above the base weight generated among two entities being second and third in the ranking of a same article. While we experimented with multiple different options, it was found that this choice allows us to maintain an interesting structure of interconnections in the graph, while lowering the noise of data and the symmetries intrinsic to our first part of the approach. The related degree distributions are also reasonable.
    \item Finally, we save the giant component of the graph.
\end{enumerate}

\begin{figure}[t] 
    \centering
    \includegraphics[width=\textwidth]{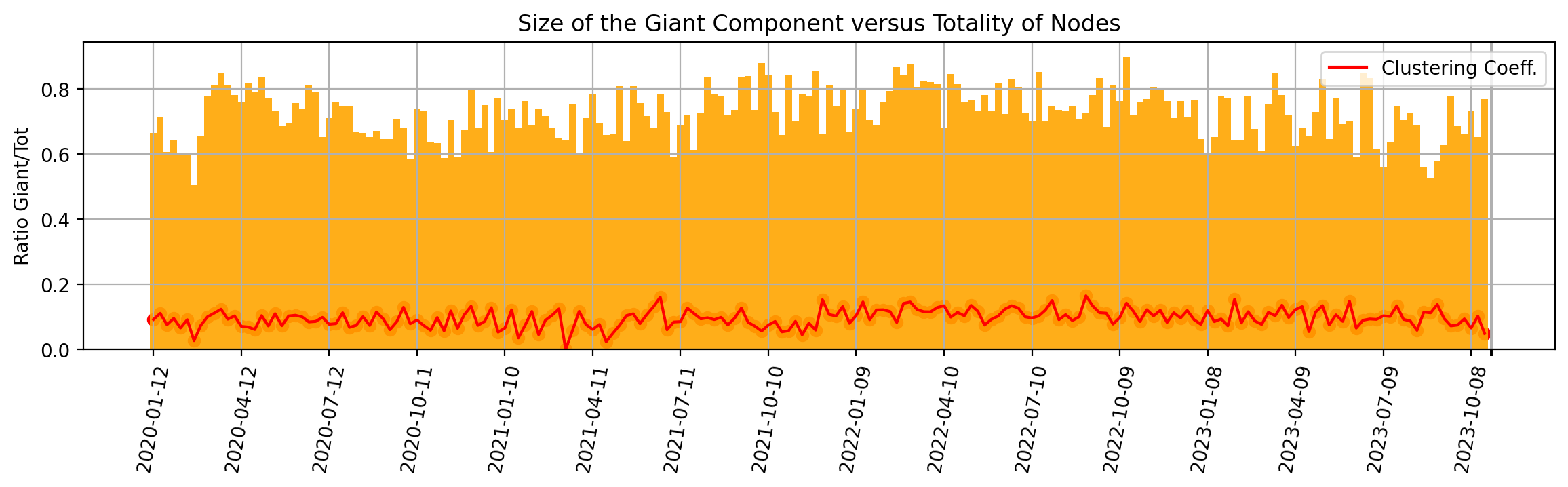}
    \caption{For each week of data, we measure the percentage of nodes that lie in the giant component of the related graph and report it in orange. In red, its average clustering coefficient is also depicted.}
    \label{fig:giant-ratio}
\end{figure}

For completeness to the above, we also show in Fig. \ref{fig:giant-ratio} both the ratio of the size of the giant component versus the total initial number of nodes and the average clustering coefficient of the network, for each week. The latter is computed as the average among the local clustering coefficient for each node in the graph, which is the proportion of the number of links between a node's neighbourhood divided by the number of links that could possibly exist between them.

The giant components tends to encompass the large majority of the nodes (and be of size $\sim 200-300$ nodes), meaning that we maintain significant amount of information, nevertheless with some intrinsic variability. When the second largest connected component has higher than average number of nodes, then it still tends to have only $\sim 20-30$ nodes. For the sake of curiosity, we looked into a few such scenarios and report here an explicative example. For the week ending on 2022-12-04, the second largest connected component of the related graph is made of the following $21$ nodes:
\begin{itemize}
    \item \textit{sam bankman-fried, us bankruptcy courts, senate agriculture committee, john j ray iii, securities and exchange commission, cryptocurrency markets, alameda research, us trustee andrew vara, ftx, commodity futures trading commission, crypto lenders and hedge funds, valar ventures, us justice department, ledger x, blockfi, monsur hussain, sec and cftc, terrausd, newyork prosecutors and sec, three arrows capital, lehman brothers}.
\end{itemize}
Clearly, this is a cluster of information related to the bankruptcy of the crypto exchange FTX in November, and its contagion to BlockFi Three Arrows Capital. However, this is disconnected from the giant component. Interestingly, we can thus inspect that e.g. the Securities and Exchange Commission (SEC) has no major other meaningful participation in the news, since otherwise the cluster would be connected to the giant component via that entity. 

Thus, our approach allows us to focus on the interconnectedness of topics via the giant component of each graph, and to consequently try to extract the main narratives among weeks and their interrelations. Other \say{large} connected components can be considered to understand disconnected topics addressed in the news, but will not be of main relevance for the \textit{global} state of the market. The average clustering coefficient of the giant component is then seen to simply oscillate around the value of $0.1$, and thus does not provide strong direct insights.
On the other hand, the degree and eigenvector centrality of nodes at different weeks will provide valuable information.

\begin{landscape}
\begin{figure}[h]
\centering
\begin{subfigure}{.9\linewidth}
    \centering
    \includegraphics[width=.99\linewidth]{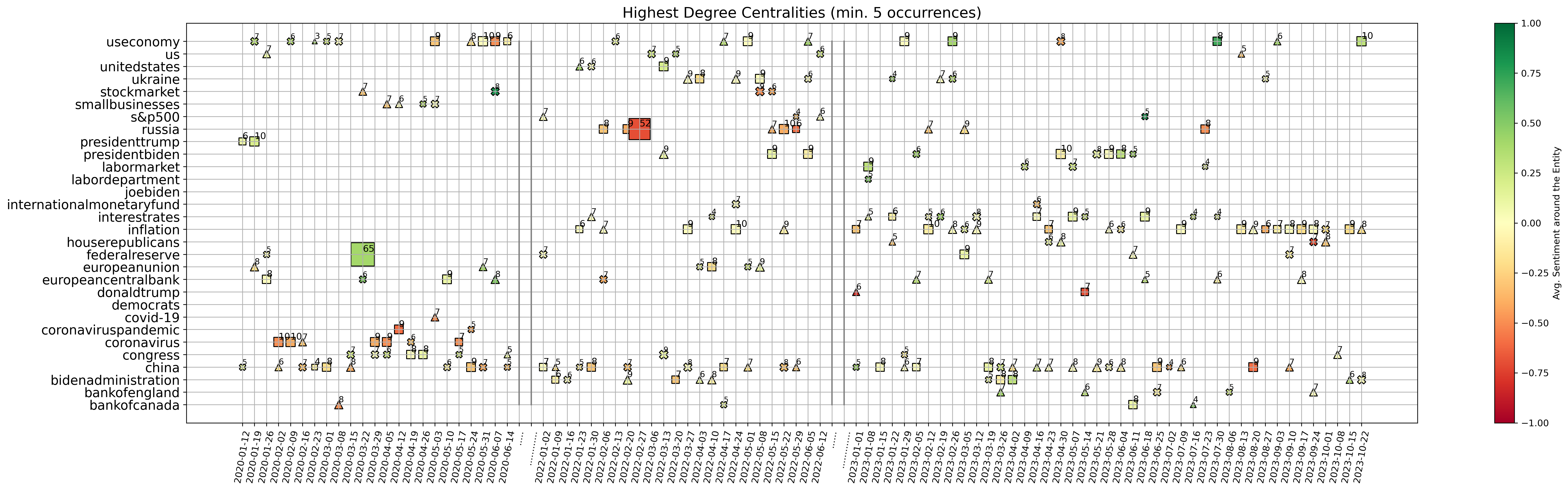}  
    \caption{Degree centralities.}
    \label{fig:figCentrA}
\end{subfigure} \\
\begin{subfigure}{.9\linewidth}
    \centering
    \includegraphics[width=.99\linewidth]{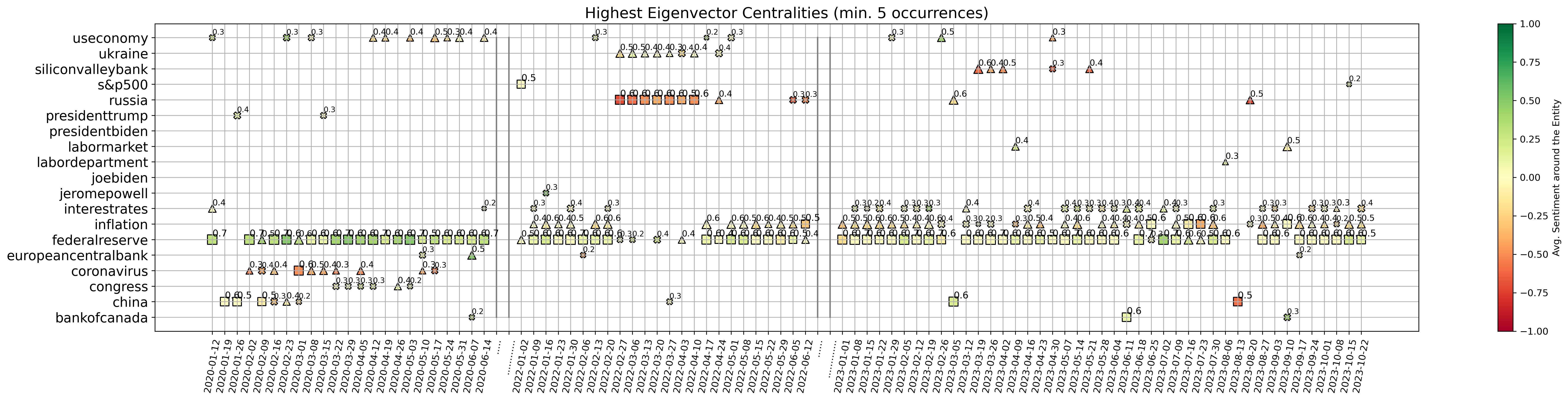}  
    \caption{Eigenvector centralities.}
    \label{fig:figCentrB}
\end{subfigure}
\caption{We show the most important entities in the graph of each week, according to two measures of node centrality, i.e. degree of the node and eigenvector centrality. For each week, we plot the 1st, 2nd, and 3rd most important entities with the shape of a square, triangle, and cross, respectively. We size and annotate each point according to the actual value of the importance measure (rounded), and colour-code it according to the average sentiment around that entity. For ease of visualisation, we show only entities that appear as highly important at least five times over our sample of weeks. In the above plots, degree centrality focuses on the size of neighbourhood of a node (i.e. the topic is addressed in connection to multiple points of view), while eigenvector centrality highlights the nodes that are close to other important nodes (i.e. measuring the interconnectedness and influence in the global environment).}
\label{fig:figCentr}
\end{figure}

\end{landscape}

\paragraph{Centralities.} We compute the weighted degree $deg(v)$ and eigenvector centrality $eig(v)$ for every node $v$ in the giant component of the graph describing each week. Then, we focus on the three nodes with highest $deg(v)$ and $eig(v)$ for the given point in time, and show them in Figs. \ref{fig:figCentrA} and \ref{fig:figCentrB}, for the two measures respectively. In particular, we focus on entities that are recurrently important, i.e. we subset to entities that have highest centrality values for at least five weeks of our sample. For each week, we then plot the first, second, and third most important entities with the shape of a square, triangle, and cross, respectively. We size and annotate each point according to the actual value of the importance measure (rounded), and colour-code it according to the average sentiment around that entity.
When we look at the nodes with highest degree, we are selecting the entities around which many news are focused. On the other hand, nodes with high eigenvector centrality are entities connected to other important entities, meaning entities lying at the center of the global net of interconnections among news.

\vfill

It is thus interesting to then analyse the different results arising from these two measures. We start from Fig. \ref{fig:figCentrB}, which shows snapshots of the evolution of nodes with highest eigenvector centrality. Clearly, the \say{Federal Reserve}, \say{inflation}, and \say{interest rates} play the major role in our net of news for the latest one year and a half. This is expected since such period is characterised by broad discussions about the persistently strong inflation, and the constant rate hikes of Central Banks. However, \say{Silicon Valley Bank} (SVB) also acquires significant importance at the end of March 2023, after it indeed collapsed on March 10th, 2023. The reason why this event has noteworthy eigenvector centrality (while it is not signalled by the degree centrality) is that SVB was the largest bank to fail since Washington Mutual closed its doors amid the financial crisis of 2008, on top of a moment of already strong fear of incoming recession in the U.S. Thus, such shock could have spread and affected current themes of discussion, meaning that our modelled interconnectedness of news was able to capture and highlight these broader concerns. Going further back in time, we see the strong concern around \say{Russia} and \say{Ukraine} at the beginning of 2022, when indeed the former invaded the latter. And clearly, \say{China} and the \say{Coronavirus} are signalled at the beginning of 2020, when the Covid-19 pandemic originated. Such entities are however not further signalled with the passing of time, meaning that the focus shifted on the consequences of such crisis rather than such topic itself.

We now move to the degree centralities shown in Fig. \ref{fig:figCentrA}. \say{Inflation}, and \say{interest rates} play again an important role, as expected. However, we have now more variability due to less interconnected topics but with high surrounding discussion. A negative shock related to \say{Russia} is clearly proposed in conjunction with the invasion of Ukraine, while a strongly positive one signals the pandemic stimulus approved by the Federal Reserve in March 2020.
Interestingly, \say{China} appears quite constantly during our full sample of data, with e.g. a point of stronger negative sentiment and concern on the week ending on 2023-08-20. This is when Evergrande filed for U.S. bankruptcy protection as China economic fears mounted, while China also unexpectedly lowered several key interest rates earlier that week. 
Finally, we can further notice that \say{President Biden} and its Administration are some times highlighted, which is sensible due to the U.S. focus of our corpus. On top of that, data show indeed that e.g. \say{Donald Trump} carries strong negative sentiment the final weeks of 2022, when the January 6 committee decided he should indeed be charged with crimes related to the assault on the U.S. Capitol happened in 2021. 



\subsection{Community detection}

The degree and eigenvector centralities of nodes in a graph can point to a concise sample of nodes of major interest, but they do not provide deeper insights on the structure and information within the totality of nodes. 
We believe that considering the problem of community detection on the proposed graphs will allow us to extract more insights on the topics characteristic of each week's news, and on the evolution of the associated narratives. We slightly distinguish between the words \say{topic} and \say{narrative}, and consider the former as the broad category or label of a series of events, while the latter as the surrounding information that can evolve over time. As an example, we would refer to the \say{Covid pandemic} as a topic, while its narrative would evolve from the initial outbreaks, to the development of vaccines, to the posterior implications of the monetary policy adopted...

\begin{figure}[h] 
    \centering
    \includegraphics[width=0.4\textwidth]{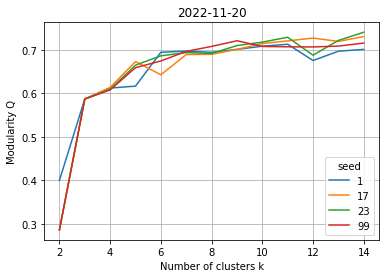}
    \caption{Looking for the optimal number of communities $k^{fuz}$ for the week 2022-11-20, to input when computing fuzzy community detection. Multiple random seeds are used to test the stability of the results against possible perturbations during NMF. A knee $k^{fuz}=5$ gets here selected.}
    \label{fig:optimal-comm-fuzzy-method}
\end{figure}

We do community detection on the graph representative of each week, both via the classic Louvain methodology and the fuzzy spectral method (Section \ref{sec:methodologyN-community}). In this way, we identify clusters of nodes highly interconnected, which we then analyse. The Louvain method automatically finds the optimal number of communities $k^{Louv}$ for each week. On the other end, we need to define the desired number $k^{fuz}$ of output communities when applying the fuzzy spectral method. To find the optimal $k^{fuz}$, we first compute the modularity $Q$ of the strict partition arising from fuzzy community detection on each given graph, for possible number of communities $k \in \{2, 3, ..., 14 \}$. Then, we automatically choose the \textit{knee} of each resulting plot, i.e. the point after which $Q$ does not significantly increase with further increases in $k$. Figure \ref{fig:optimal-comm-fuzzy-method} shows an example of such plot, for the week ending on 2022-11-20. Since the result is based on NMF, we always compute the clustering from multiple random seeds, and then take the mode of the suggested knees (despite the trends tend to be uniform, as the plot hints). Of course, we also investigate the diffusion parameter $\beta$ in parallel, and fix it to $\beta = 1$. Then, we save the stability index of each node for the optimal partition of each graph.

For our set of graphs of interest, the average $k^{Louv}$ is $17$ (with standard deviation $\pm 3$), while the average $k^{fuz}$ is $5$ (with standard deviation $\pm 1$). This means that the diffusion kernel employed in the fuzzy spectral method, which captures long-range relationships between nodes, allows us to achieve more tractable clusterings. These also encapsulate broader structures within the networks, and will be preferred against Louvain communities.
For the sake of completeness, we compute the Adjusted Rank Index (ARI) among the clusterings suggested by the two methodologies, for each week. The ARI is a similarity measure between two clusterings, which is computed by considering all pairs of samples of data, and counting pairs that are assigned in the same or different clusters in the two tested clusterings. If the clusterings are identical, the ARI reaches its maximum value of $+1$, while this measure drops up to $-0.5$ for especially discordant clusterings. The ARI between partitions generated by our Louvain and fuzzy methodologies is $0.37\pm 0.11$ on average, meaning that there is a good level of agreement between the results, and increasing our confidence in focusing on the fuzzy spectral methodology from this point onward.

Importantly, we have been considering the strict partition generated by the introduced spectral method so far, but have not yet focused on its fuzziness component and stability of nodes. This is indeed leveraged upon now, since our hypothesis is that the communities highlighted by such method can be directly mapped to the main topics discussed within news. Thus, dropping the most unstable nodes will allow us to achieve a clearer and cleaner view. For each week, we proceed as follows:
\begin{enumerate}
    \item For each community identified, we extract the list of nodes belonging to it.
    \item For each related node $i$, we check whether $S_i \leq 2$ (i.e. if the level of membership in the most likely community is less than twice the level of membership to the second one). If the relationship is satisfied, then we label the node as \textit{unstable}. Importantly, such threshold is chosen by looking at the distribution of stability indices over nodes for a sample of plots.
    \item Then, we take all the news published that week. For each one node, we consider all articles for which the node is one of its five main entities. We further check whether all the three main entities of the article are in the nodes of the community under investigation, but not in the associated set of unstable nodes. The articles that satisfy all conditions are kept.
    \item For each final set of articles representative of each community of the week, we save the related information (i.e. GPT-generated summary and GPT-generated abstract).
\end{enumerate}

\begin{figure}[t]
\centering
\begin{subfigure}{.99\linewidth}
    \centering
    \includegraphics[width=.99\linewidth]{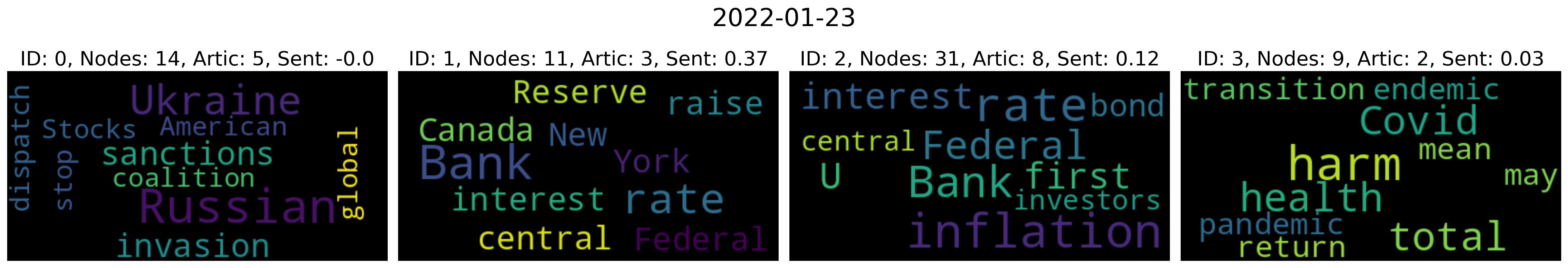}  
    \caption{Word-clouds representative of the topics identified for the week ending on 2022-01-23.}
    \label{fig:WC1}
\end{subfigure} \\
\begin{subfigure}{.99\linewidth}
    \centering
    \includegraphics[width=.99\linewidth]{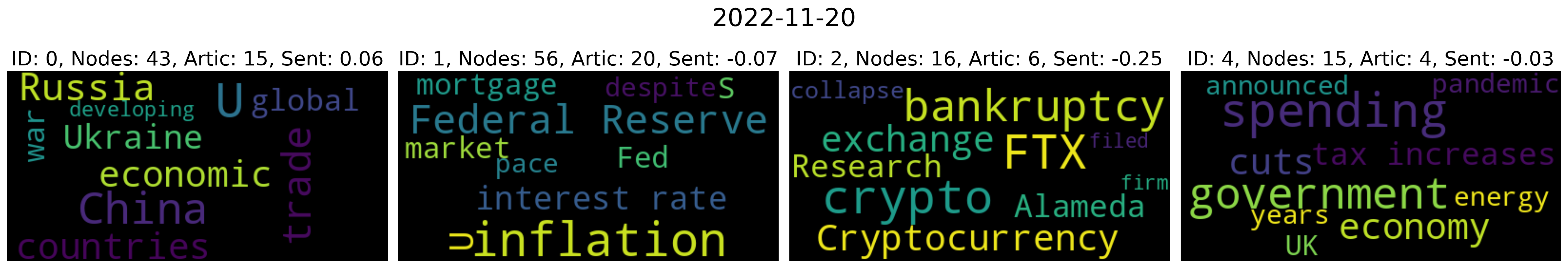}  
    \caption{Word-clouds representative of the topics identified for the week ending on 2022-11-20.}
    \label{fig:WC2}
\end{subfigure} \\
\begin{subfigure}{.99\linewidth}
    \centering
    \includegraphics[width=.99\linewidth]{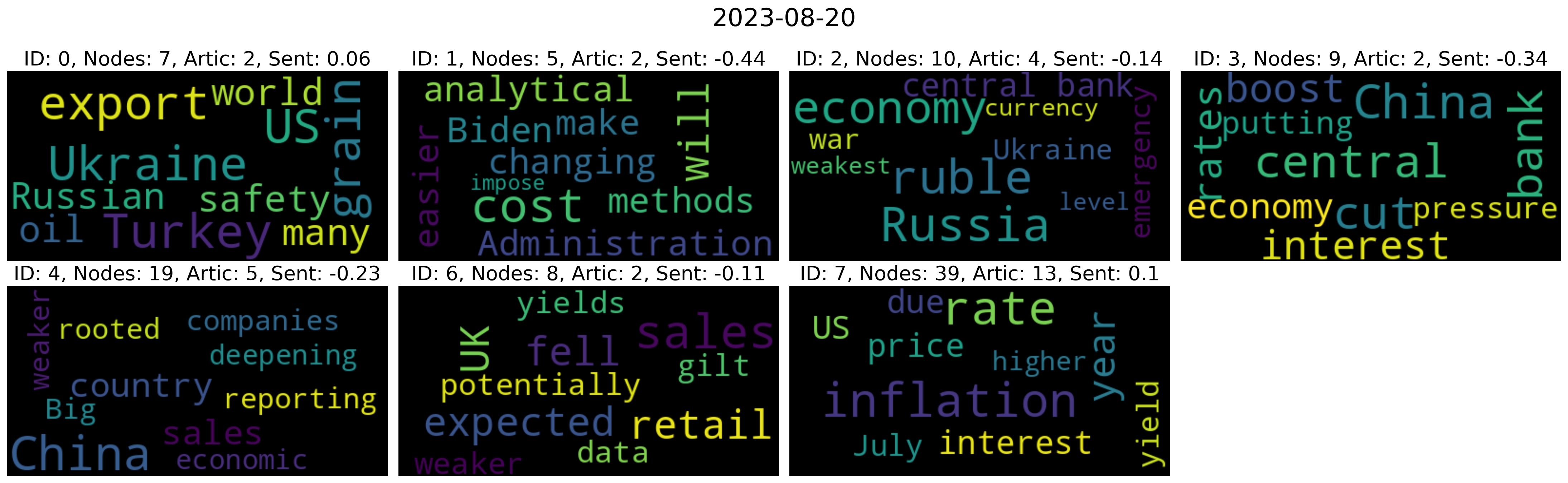}  
    \caption{Word-clouds representative of the topics identified for the week ending on 2023-08-20.}
    \label{fig:WC3}
\end{subfigure} \\
\caption{Word-clouds for the topics within three sample weeks of news. If no articles are seen to belong to a topic-community after our stability analyses, then we drop the related empty word-cloud. This is the reason why some IDs are missing in the above plots.}
\label{fig:wordclouds}
\end{figure}

To summarise the above, we construct a set of highly representative articles for each community identified within the graph describing each week of data. We are very stringent with our stability requirements, and indeed drop on average $16$ articles per community. However, this allows us to find the articles at the heart of each community, which enable the most effective downstream analyses.
We then concatenate all related GPT summaries (and abstracts) for each community, and visualise them via word-clouds\footnote{A word-cloud is a collection of words depicted in different sizes and strengths to indicate the frequency and importance of such word in the text.} to see whether they produce coherent topics. In parallel, we also feed such joint data into GPT and ask it to produce a newly associated summary, which should then allow us to easily interpret each topic (and narrative) of the week. The examples proposed in Fig. \ref{fig:wordclouds} aim to provide initial evidence of the accomplishments achieved by our methodology, where we show word-clouds of topics found for the weeks ending on 2022-01-23, 2022-11-20, and 2023-08-20. The related summaries of joint summaries are reported in Tables \ref{tab:summ-of-summ-2022-01-23}, \ref{tab:summ-of-summ-2022-11-20}, and \ref{tab:summ-of-summ-2023-08-20}, respectively, and we just mention that summaries on abstracts are highly similar. Importantly, we remark that if no articles belong to a community after our stability analyses, then clearly neither a word-cloud is shown nor a summary reported.

By studying the generated topics via word-clouds and proposed summaries, we find interpretable results that accurately map to the broad major events described by the data for each related week. Clusters for week 2022-11-20 are a nice example. On the other hand, we also see ability to extract the precursor worries of a Russian invasion of Ukraine from the findings of week 2022-01-23. This could be of great use for a thorough and systematic study of early signs of crises with more historical data. Moreover, one could compare the time of appearance of such signs across results computed from different corpora, e.g. collected by news from Journals of different countries. Finally, we show the clusters for week 2023-08-20, since this is one of our latest occurrences of a market dislocation event (refer to Fig. \ref{fig:vol-indices}). Interestingly, we see that many clusters are identified depicting concerns across themes of discussion. These different topics still need to be part of the giant component of our graph, implying that we have an instance of market turbulence related to high entropy of discussions within news that must be somehow more broadly interconnected.

\begin{table}[h]
    \centering
    \begin{tabular}{|c|p{15cm}|}
    \hline
       ID  & GPT Summary of Summaries \\
       \hline
        0 & China's stocks rose and Hong Kong's stocks fell after the government cut interest rates, while the US is preparing financial sanctions on pro-Russian agents in Ukraine to deter Russia from invading its neighbor. \\
        1 & US and Canada are considering raising interest rates, while China has cut its benchmark lending rates to support its economy, and the Federal Reserve Bank of New York has appointed a new official to oversee financial-market operations. \\
        2 & Tax revenue is up, inflation is rising, central banks have differing views on interest rates, and US companies may struggle to generate high profits in the coming year. \\
        3 & The transition of Covid-19 from pandemic to endemic could still have lasting effects on health and the economy, and policies should aim to minimize harm from both infection and prevention measures. \\
    \hline
    \end{tabular}
    \caption{GPT summaries on the joint summaries of articles belonging to each community for week ending on 2022-01-23, after our stability filtering.}
    \label{tab:summ-of-summ-2022-01-23}
\end{table}

\begin{table}
    \centering
    \begin{tabular}{|c|p{15cm}|}
    \hline
       ID  & GPT Summary of Summaries \\
       \hline
        0 & The G-20 summit addressed global economic challenges, including the war in Ukraine and the weaponization of food production, while the US imposed sanctions on individuals and companies linked to Russia, and new intelligence showed that parts in Iranian drones downed in Ukraine were manufactured by companies in allied nations. \\
        1 & Investors are optimistic about decreasing inflation, but earnings are becoming a threat as Wall Street analysts slash profit forecasts, while the Federal Reserve warns of the need to raise interest rates to control inflation, and the housing market cools down due to a lack of buyers for mortgage bonds. \\
        2 & Cryptocurrency firm FTX and its investment arm Alameda Research have filed for bankruptcy, leaving customers facing potential losses in an unregulated sector, with details still scarce. \\
        3 & NA \\
        4 & The UK government has announced tax increases and spending cuts to reduce government debt relative to the economy, becoming the first major Western economy to limit spending growth after years of fiscal stimulus during the pandemic and recent energy subsidies. \\
    \hline
    \end{tabular}
    \caption{GPT summaries on the joint summaries of articles belonging to each community for week ending on 2022-11-20, after our stability filtering.}
    \label{tab:summ-of-summ-2022-11-20}
\end{table}

\begin{table}
    \centering
    \begin{tabular}{|c|p{15cm}|}
    \hline
       ID  & GPT Summary of Summaries \\
       \hline
        0 & The US is in talks to increase alternative export routes for Ukrainian grain after Russia pulled out of an agreement, with a US-backed plan involving increasing capacity for Ukraine to export via the Danube River, while concerns have been raised over the safety and environmental impact of Turkey-based Beks Ship Management's transportation of Russian oil. \\
        1 & The Biden Administration is changing its analytical methods to make it easier to impose new rules while disguising their cost, resulting in increased regulatory costs on the economy. \\
        2 & Russia's ruble has fallen to its weakest level in over a year due to Western sanctions and the war in Ukraine, prompting an emergency meeting by the central bank, but analysts suggest that factors driving down the currency are largely out of their control. \\
        3 & China's interest rate cut to boost the economy is putting pressure on the currency and causing declines in major stock indexes. \\
        4 & Big American companies rooted in China are reporting weaker sales and turning to other countries for imports due to China's deepening economic slump. \\
        5 & NA \\
        6 & UK retail sales fell more than expected in July due to bad weather and economic issues, causing UK gilt yields to drop and potentially reducing the need for Bank of England interest-rate rises to combat inflation. \\
        7 & Inflation is decreasing in the US, but rising energy and food prices may cause turbulence, while Canada's inflation rose unexpectedly and may lead to a rate increase, and Latin American central banks are cutting interest rates as inflation eases. \\
    \hline
    \end{tabular}
    \caption{GPT summaries on the joint summaries of articles belonging to each community for week ending on 2023-08-20, after our stability filtering.}
    \label{tab:summ-of-summ-2023-08-20}
\end{table}

\begin{figure}[h]
\centering
\begin{subfigure}{.99\textwidth}
    \centering
    \includegraphics[width=.99\linewidth]{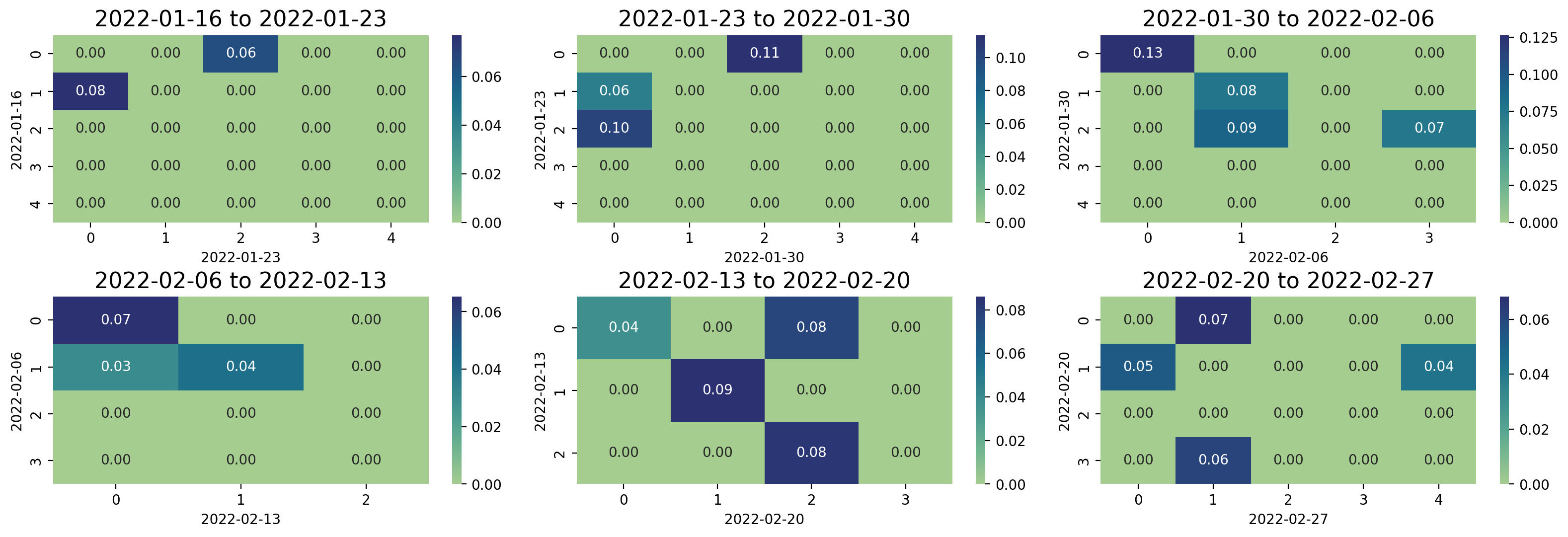}  
    \caption{Matrices of Jaccard similarity between sets of nodes defining communities for consecutive weeks over time. In this way, we investigate potential links for topics (and narratives) over time.}
    \label{fig:follow-keyword}
\end{subfigure}\vspace{0.5cm} \\
\begin{subfigure}{.5\textwidth}
    \centering
    \includegraphics[width=\linewidth]{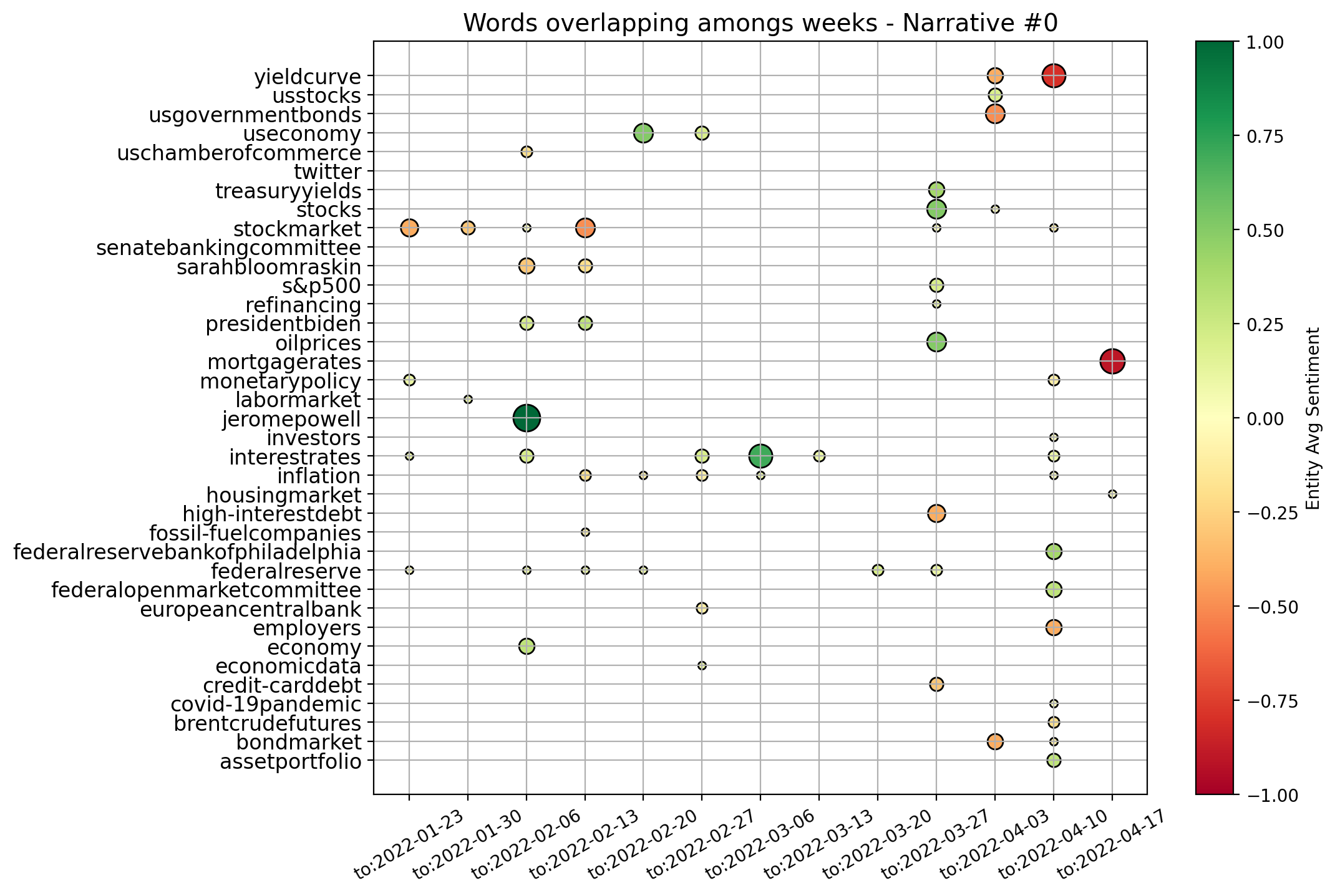}  
    \caption{Tracking the narrative with ID 0 on 2022-01-16.}
    \label{fig:follow-jaccard-0}
\end{subfigure} \hfill
\begin{subfigure}{.48\textwidth}
    \centering
    \includegraphics[width=\linewidth]{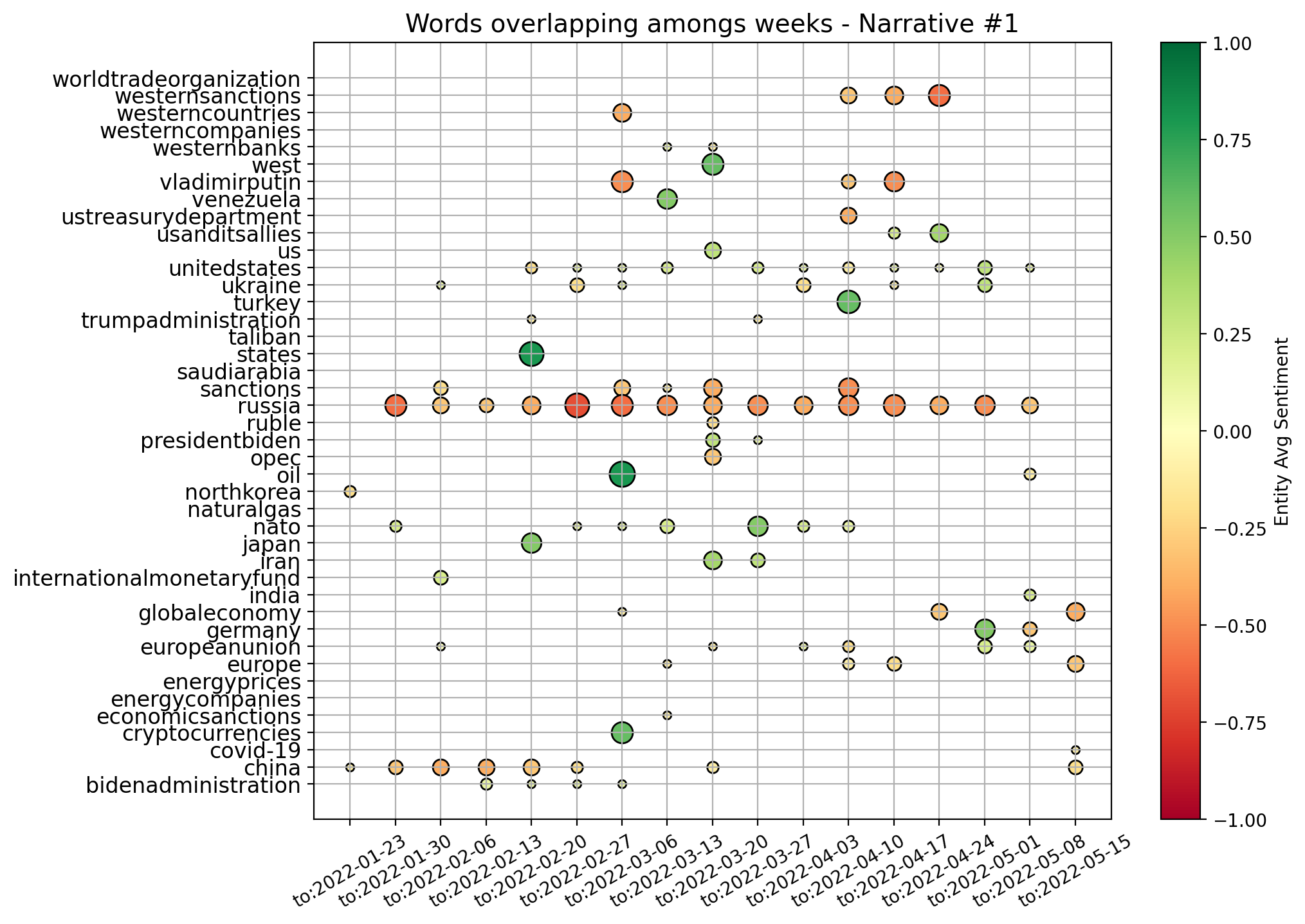}  
    \caption{Tracking the narrative with ID 1 on 2022-01-16.}
    \label{fig:follow-jaccard-1}
\end{subfigure}
\caption{We test whether we can track interpretable narratives over time, via the similarity of communities of nodes found for each two consecutive weeks. Narrative 0 is seen to relate to inflation and interest rates concerns, while Narrative 1 focuses on Russia.}
\label{fig:Jaccards-over-time}
\end{figure}

\paragraph{Evolution of narratives.} For each week of data, we found an optimised partition of its graph's nodes into communities with interpretable related economic themes. This is however a static representation of topics, which we now try to extend by linking and tracking such clusters over time. Thus, we consider each pair of consecutive weeks, and compute the Jaccard Index $J(C_{i,1}, C_{j,2})$,
\begin{equation}
    J(C_{i,1}, C_{j,2}) = \frac{|C_{i,1} \cap C_{j,2}|}{|C_{i,1} \cup C_{j,2}|},
\end{equation}
for any pair of respective communities that identify topics. We call $C_{i,1}$ and $C_{j,2}$ the sets of nodes that belong to any one community identified within the first and second week, respectively. The Jaccard Index allows us to compute a measure of similarity between communities from the overlap of their nodes, meaning that we can assess which clusters (and so topics) are the most similar over consecutive weeks and link them. We also restrict our results to have $|C_{i,1} \cap C_{j,2}| > 1$, to increase the associated significance. Figure \ref{fig:follow-keyword} shows a sample of matrices that link topics from their Jaccard similarity, for a series of pairs of consecutive weeks. The simplest way to track a narrative is thus to start from its first acknowledged community, and \say{jump} to communities at consequent points in time by following the series of highest similarity scores. In our case, we resolve multiple mappings among communities by considering only the maximum entry per column and row in the proposed similarity matrices, but this can be of course improved.

\begin{figure}[t] 
    \centering
    \includegraphics[width=\textwidth]{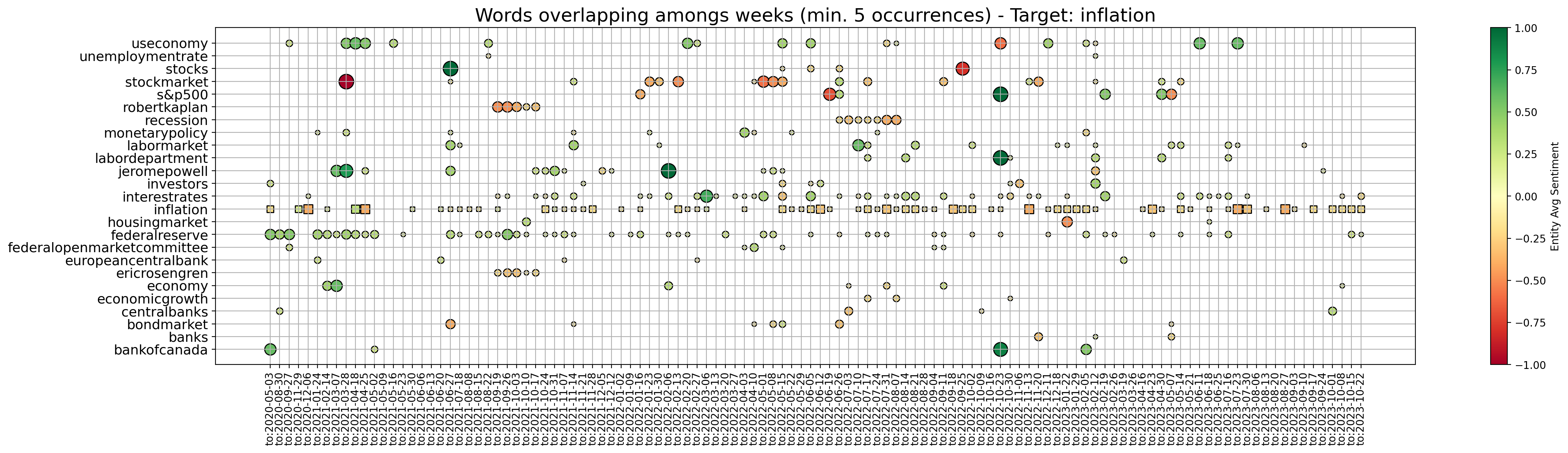}
    \caption{We define inflation as a keyword of interest, and track the recurrent entities belonging to its same community for consecutive weeks over time.}
    \label{fig:inflation-tracking}
\end{figure}

As an illustrative case, community $0$ in Fig. \ref{fig:follow-keyword} at week 2022-01-16 is most similar to community $2$ at week 2022-01-23, which is then most similar to community $0$ at week 2022-01-30, and the latter to community $0$ again at week 2022-02-06, and so on. Of course, if no mapping is at some point found, then the narrative \say{has broken}. In Figures \ref{fig:follow-jaccard-0} and \ref{fig:follow-jaccard-1}, we respectively track both the shown topics $0$ and $1$ with the simple methodology proposed. We plot the nodes that are found to overlap for the most similar communities between each two consecutive weeks, until a mapping persists. It is direct to see that narrative $0$ relates to inflation, interest rates, and the behaviour of Central Banks, while narrative $1$ concerns Russia and the themes surrounding its war with Ukraine.
In parallel, we also deploy a slightly different approach and just save the communities to which a chosen keyword belongs over time. In this way, we try to better investigate the surrounding evolving narrative. As an example, Fig. \ref{fig:inflation-tracking} shows the overlap of entities (generally plotted as dots) belonging to the same communities as the word \say{inflation} (plotted as squares) over consecutive weeks. New characters are highlighted, such as Eric Rosengren and Rober Kaplan, and an important period of concern is signalled around July 2022 and the word \say{recession}. Indeed, concerns about inflation, interest rates, and the fall of Gross Domestic Product (GDP) in the first two quarters of 2022, led to heightening recession fears during the month of July.

\vfill

\subsection{News and financial markets dislocations}

\begin{figure}[b]
    \centering
    \includegraphics[width=\linewidth]{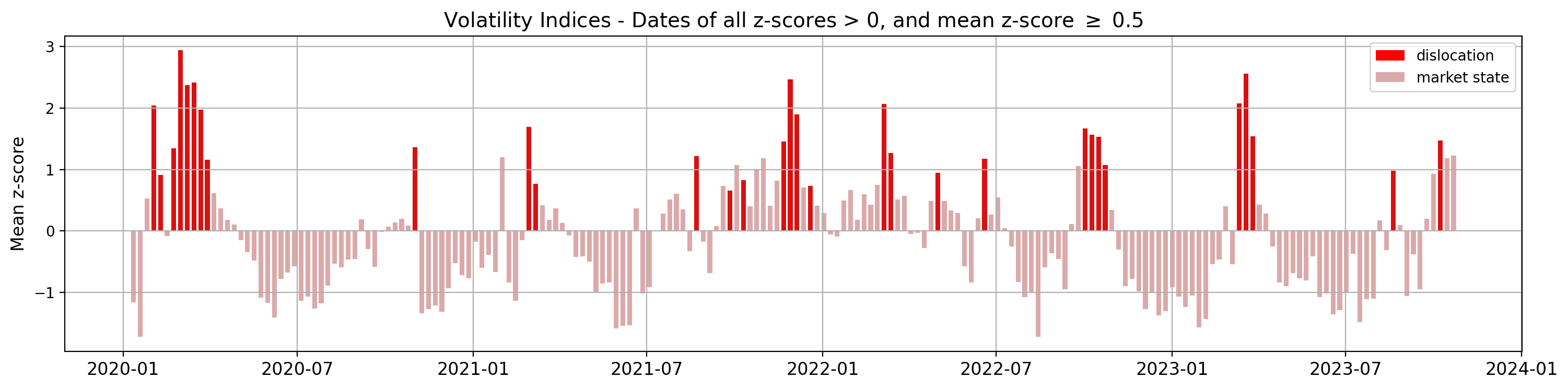}
    \caption{In bright red, we depict dates in which all our four z-scores are strictly positive, and their mean is also above $0.5$. These are our dislocation dates of reference (to which we assign label $+1$). On the other hand, the simple average on the four z-scores is plotted in faded red. It is clear that dislocation dates are an under-sampled category, which we expand via the SMOTE technique.}
    \label{fig:binary-target-z-scores}
\end{figure}

\textcolor{black}{Thanks to the analyses just completed, we have supporting evidence on the ability of our graph constructs to advance in the task of topic detection and narrative characterisation. However, we now desire to test whether information on such structure of news allows us to unravel novel insights on broad market dislocations.} As introduced in Section \ref{sec:z-scores-dislocations}, we describe the state of the market by the z-scores of our volatility indices. Here, we then define a market dislocation as a week when all four of our z-scores are strictly positive, and their average is above $0.5$. Such weeks are identified with a $+1$ label, while all other periods with a $0$ label, implying that we now have a binary variable to use as target of a logistic regression model (see Section \ref{narr:logistic-regression}). For the sake of clarity, Fig. \ref{fig:binary-target-z-scores} shows our points of interest.

\begin{figure}[t]
    \centering
    \includegraphics[width=0.6\linewidth]{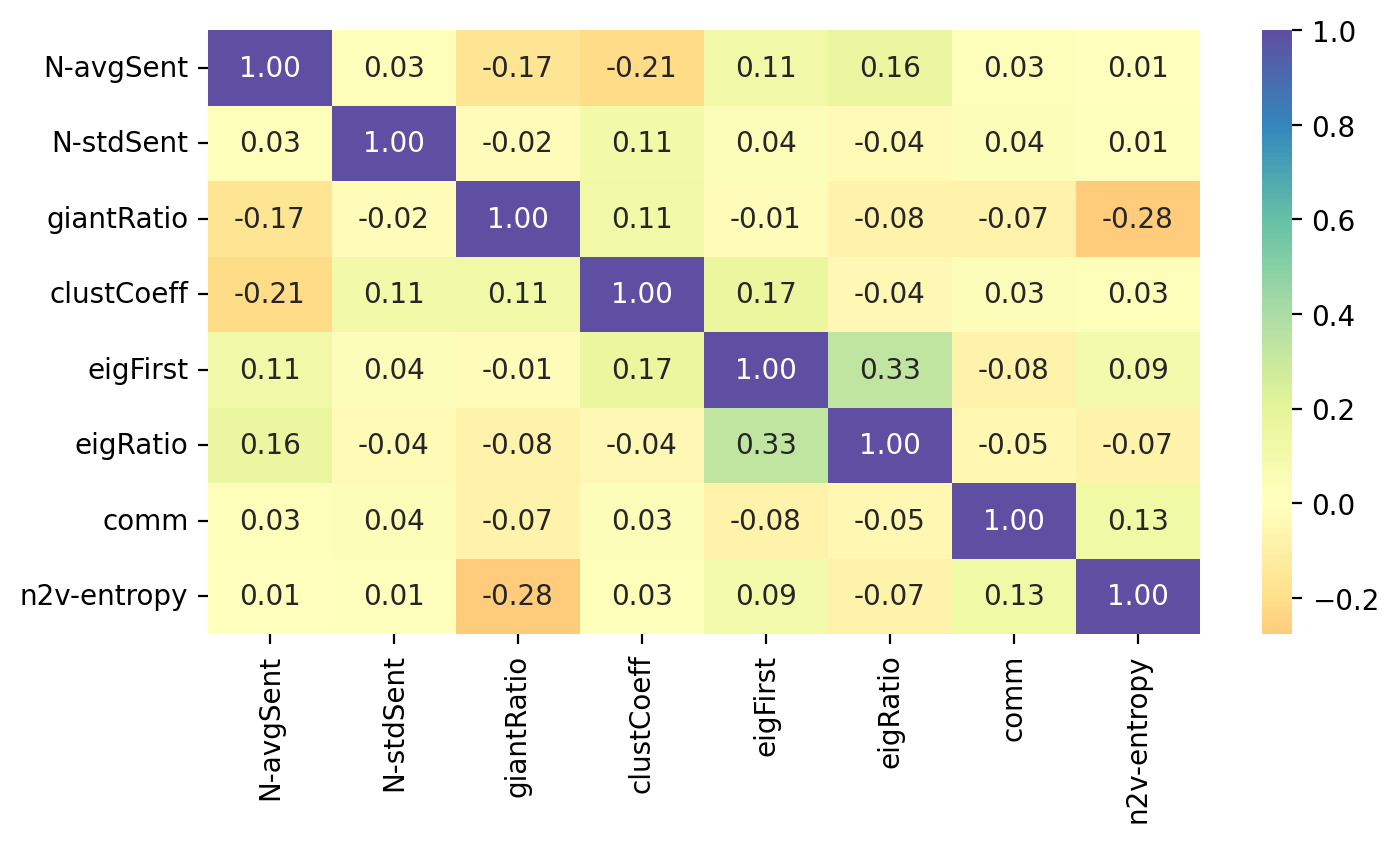}
    \caption{Correlation matrix of the features tested in the logistic regression, but without including market features.}
    \label{fig:corr-feat-narr}
\end{figure}

To run the suggested logistic regression, we need to build a set of features that characterise each week. Then, we can either map them to contemporaneous or next states of the market, to respectively investigate relationships between news and either unfolding or incoming dislocations. The features we consider can be seen to belong to three different categories, namely:
\begin{enumerate}
    \item \textit{Market features}. Features on the current state of the market (used in the case of predictive logistic regression, and as a benchmark). These are the current average of our four z-scores of volatility indices, and the difference of such value from the mean of the previous week.
    \item \textit{News features}. Features based on the raw corpus of news available.
    \item \textit{Graph(+) features}. Features extracted from the set of graphs that we built to capture the structure and interconnectedness of news within weeks. We further leverage on node2vec (n2v) embedding methodology.
\end{enumerate}
Due to the limited number of data points (i.e. weeks) available, we build and test only features that we believe to be the most significant and meaningful for each category. Indeed, it is a common proxy to allow at least $\sim 20$ outcomes for each independent variable tested. Also, we check the correlation among each pair of variables, to drop such features that would cause problems of multicollinearity and invalidate any results of the regression. 
Table \ref{tab:features} summarises our final set of $10$ tested features and chosen related tags, while Fig. \ref{fig:corr-feat-narr} shows their correlation matrix from Pearson's test. As desired, all the retrieved features show either small or negligible correlation among each other. 

Before proceeding to the logistic regression model itself, two points need to be now better discussed. These are the further features initially contrived but then dropped, and the methodology designed to construct the proposed \say{n2v-entropy} feature (i.e the last row in our table of features, with a \say{Graph+} class label). 
Our \textit{news features} limit themselves to measures of the sentiment across articles, since we can safely assume that sentiment is indeed one of the most important market-related features that can be extracted from a plain corpus of economic news. We could have further included e.g. the specific number of news available for each week, but this would have been a noisy metric due to its strong dependence on the data collection step of our framework. Importantly, allowing for such simple but meaningful features of news gives us the opportunity to both test whether news have significant relationships with market dislocations, but also to assess whether our proposed graph representation of news is useful at all.
Then, our \textit{graph features} focus on characteristics that we can directly compute and extract from our graph representation of news in a week. For similar reasons as above, we do not include the absolute number of nodes and highest degree value of a graph (while we do compute the ratio between the size of the giant component and total number of nodes). Then, we do consider the average clustering coefficient of the network, but drop the average degree due to related high correlation. Due to a similar instance of high correlation, we unsurprisingly also drop the average and standard deviation of the sentiment of nodes in the graph. On the other hand, we retain the highest absolute value of nodes' eigenvector centrality, and the ratio between such first and second highest values. The latter features allow us to better account for the structure of the network, and the potential presence of highly influential hubs. Finally, we also add to our set of features the optimal number of communities to partition the graph into, as identified by the fuzzy community detection methodology.

\begin{table}[b]
    \centering
    \begin{tabular}{|c|c|p{12cm}|}
    \hline
       \textbf{Class}  & \textbf{Name} & \textbf{Description} \\
       \hline
        Market & z-vols & Average of our four volatility indices' z-scores (i.e. VIX, MOVE, VIX FX, and MRI).  \\
        Market & z-volsD & Difference of average z-scores between the current week and the previous one. \\
        \hdashline
        News & N-avgSent & Average sentiment of news in a week, where each article's sentiment is computed as the weighted mean of entities' sentiment. \\
        News & N-stdSent & Standard deviation of the above-mentioned sentiment of news in a week. \\
        \hdashline
        Graph & giantRatio & Ratio between the size of the giant component and the total number of possible nodes.\\
        Graph & clustCoeff & Average clustering coefficient of the graph. \\
        Graph & eigFirst & Highest value of eigenvector centrality of nodes in the graph. \\
        Graph & eigRatio & Ratio between the first and second highest eigenvector centralities. \\
        Graph & comm & Optimal number of communities as identified from fuzzy community detection.\\
        \hdashline
        Graph+ & n2v-entropy & Proxy for the total entropy of node2vec embeddings lying in their multi-dimensional space, by leveraging on the Kullback–Leibler divergence measure. \\
    \hline
    \end{tabular}
    \caption{Final set of features evaluated within our logistic regression models. These features are tested with the aim to unravel relationships between news and financial market dislocations.}
    \label{tab:features}
\end{table}

Regarding our \textit{graph+ feature}, this is based on encoding nodes of each graph into an embedding space via the node2vec methodology (see Section \ref{methodology-narratives-embeddings}). Node2vec has been shown to perform significantly well across a variety of practical tasks (such as mapping accuracy, greedy routing, and link prediction) on real-world graphs having from dozens to thousands of nodes \cite{PhysRevE.104.044315}. It also suits us, since in this way we can design a metric that accounts for an holistic view of the topology of each graph. Importantly, node2vec embedding spaces cannot be compared across graph, meaning that we need to construct an ad-hoc downstream measure to characterise each point in time. 

First of all, we begin by testing embeddings for multiple combinations of node2vec hyperparameters, in order to gauge the associated sensitivity of results. Similarly, we experiment with different combinations of the return parameter $p_{emb}$ and the in-out one $q_{emb}$, in order to investigate the sampling strategy to adopt.
Embeddings are thus computed for all (i.e. $24$) the possible combinations of:
\begin{itemize}
    \item output vectors dimensions $\in \{8, 16\}$,
    \item length of simulated walks $\in \{10, 20\}$,
    \item number of walks per node $\in \{10, 20\}$,
    \item $(p_{emb},q_{emb}) \in \{(1,1), (4,2), (4,0.5)\}$.
\end{itemize}
After extensive exploratory data analysis, we choose the following final combination of (hyper)parameters for our embeddings: output vectors dimension $8$, length of simulated walks $20$, number of walks per node $20$, and $(p_{emb},q_{emb})=(1,1)$. Overall, we saw higher stability of outputs for such values of length of walks and number of walks per node, while favouring larger dimensions does not significantly affect results. Finally, we choose $(p_{emb},q_{emb})=(1,1)$, to allow the weight of edges to directly dominate the sampling strategy, i.e. the probability of transition to a neighbouring node.
Thanks to the final set of embeddings achieved, we can then proceed to design related features that alternatively describe each week's structure of news. One in particular is here proposed. 

For each week, we have a related graph, and a consequent embedding for the nodes of such graph in a $8$-dimensional space. We then design the following proxy for the overall entropy of nodes in the identified embedding space:
\begin{enumerate}
    \item For each (orthogonal) dimension, we approximate the distribution of data on such axis by computing the probability of occurrence of points in bins of width $0.1$.
    \item We then calculate the Kullback–Leibler divergence $D_{KL}$ of such distribution $P_{data}$ from the uniform probability distribution $P_{unif}$. This is
    \begin{equation}
        D_{KL}(P_{data} || P_{unif}) = \sum_{b \in B} P_{data}(b) \times \ln \Big( \frac{P_{data}(b)}{P_{unif}(b)} \Big),
    \end{equation}
    where $B$ is the set of possible outcomes (i.e. bins).
    \item Finally, we average these divergences computed on our eight dimensions to achieve a proxy of global entropy. This is indeed named \say{n2v-entropy}.
\end{enumerate}

\paragraph{Contemporaneous and predictive analyses.} Now that we have introduced our target variable and motivated the features to test, we proceed to the implementation of our logistic regression models. Importantly, we consider data only from June 1st, 2020 to avoid dislocations related to the Covid-19 crisis. Such extreme exogenous shock would indeed prevent us to focus on more subtle and isolated moments of dislocations, which we believe to be the ones in need of better understanding. We also complete only in-sample tests, due to both the limited number of data points available and our specific interest on assessing whether useful information lies within the modelled \textit{structure and interconnectedness} of news, on which one can then build upon if successfully proven. Since our dislocation dates are strongly under-sampled, we leverage on the SMOTE technique introduced in Section \ref{narr:logistic-regression} to equally re-balance our target classes. We also standardise all features, and then train a logistic regression model for the following two cases:
\begin{enumerate}
    \item We allow all features but \textit{market} ones, and try to unravel connections between news and a contemporaneous state of market dislocation.
    \item We allow all features, and try to predict an incoming state of market dislocation.
\end{enumerate}
Despite prediction of incoming market dislocations is the most desirable end goal, we are indeed strongly interested in unravelling the features of news' structure that are connected to such critical weeks. These would increase our understanding of such events with statistical confidence, and could suggest novel research ideas.

For each one of the two above cases, we do Recursive Feature Elimination (RFE) to complete feature selection, and require each chosen attribute to be significant at least at the $p$-value $<0.05$ level. The full specifications of our final models are reported in Tables \ref{table-model-latex-contemp} and \ref{table-model-latex-pred}, for our contemporaneous and predictive cases, respectively. Similarly, Tables \ref{confusion-matrix-contemp} and \ref{confusion-matrix-pred} report the associated confusion matrices on their left, which arise from testing back our resultant models on the (not over-sampled) initial data. The latter tables also report a few further metrics on their right. Indeed, we show the precision of our classifier, which is its ability to not label a sample as positive if it is negative (i.e. the ratio between true positives divided by the sum of true positives and false positives). Then, the recall assesses the ability of the classifier to find all the positive samples (and is computed as the ratio between true positives divided by the sum of true positives and false negatives).
Finally, the F-beta score can be interpreted as a weighted harmonic mean of the precision and recall, and reaches its best value at 1 and worst score at 0.

Our model for the contemporaneous relationship between news and market dislocations proposes five significant features. Both the average and standard deviation of simple news' sentiment (i.e. \say{N-avgSent} and \say{N-stdSent}) are among them, with the former having unsurprisingly negative coefficient. Thus, a more positive sentiment of news clearly lowers the probabilities of being in a moment of market dislocation. We also believe that a higher standard deviation generally implies a subset of articles with stronger negative sentiment rather than positive, and thus a positive coefficient is seen. The \say{giantRatio}, \say{eigFirst}, and \say{comm} features are instead related to our proposed graph characterisation of weekly news, and have all positive coefficient. Having a larger proportion of nodes kept within each graph's giant component implies that articles are more interconnected within each other, but a higher number of communities suggests that there are also more themes of discussion. Then, a larger first eigenvector centrality value suggests very influential hubs (i.e. concepts) within the construct. Merging all such information, we can see that the model hints to a relationship between market dislocations and high entropy of discussion among news. The latter need anyways to be in some way interconnected by construction, and thus point to contagion effects among themes of concern.

On the other hand, our model to predict incoming market dislocations proposes only three significant features. Clearly, the current state of the market is the strongest predictor for possible dislocations, but both \say{eigRatio} and \say{n2v-entropy} are still found significant and with not-negligible participation in the definition of the outcome probability. Such attributes have coefficients with negative sign, for which we now propose a motivation. A lower ratio between the first and second eigenvector centralities implies that the two related nodes have importance values more similar in magnitude. Thus, this refers to a situation in which there are competing hubs of importance, and likely competing points of concerns. Similarly, a lower Kullback-Leibler divergence actually implies higher entropy within the generated node2vec nodes' embeddings. Such higher entropy points to less uniform structure within narratives, which seems to indeed encode an early alert of incoming market dislocations.

Despite being very simple initial analyses, the considerations just outlined provide a baseline of evidence for a connection between news structure (i.e. beyond their mere sentiment) and market dislocations. Therefore, these first results further motivate the analyses proposed at the beginning of this project, and should prompt more studies that leverage on our graph construct to investigate broad corpora of news.

\begin{table}[h]
\caption{Assessment of the \textit{contemporaneous} relationships between news and moments of market dislocations.}
\label{table-model-latex-contemp}
\begin{center}
\begin{tabular}{llll}
\hline
Model:              & Logit            & Pseudo R-squared: & 0.212       \\
Dependent Variable: & y                & AIC:              & 342.0211    \\
Date:               & 2023-11-04 13:33 & BIC:              & 360.6062    \\
No. Observations:   & 304              & Log-Likelihood:   & -166.01     \\
Df Model:           & 4                & LL-Null:          & -210.72     \\
Df Residuals:       & 299              & LLR p-value:      & 1.7552e-18  \\
Converged:          & 1.0000           & Scale:            & 1.0000      \\
No. Iterations:     & 19.0000          &                   &             \\
\hline
\end{tabular}
\end{center}

\begin{center}
\begin{tabular}{lrrrrrr}
\hline
           &   Coef. & Std.Err. &       z & P$> |$z$|$ &  [0.025 &  0.975]  \\
\hline
N-avgSent  & -0.4409 &   0.1463 & -3.0130 &      0.0026 & -0.7277 & -0.1541  \\
N-stdSent  &  0.6774 &   0.1516 &  4.4672 &      0.0000 &  0.3802 &  0.9746  \\
giantRatio &  0.4679 &   0.1452 &  3.2223 &      0.0013 &  0.1833 &  0.7525  \\
eigFirst   &  0.4502 &   0.1556 &  2.8932 &      0.0038 &  0.1452 &  0.7551  \\
comm       &  0.4539 &   0.1378 &  3.2934 &      0.0010 &  0.1838 &  0.7240  \\
\hline
\end{tabular}
\end{center}
\end{table}

\vfill

\begin{table}[h]
\caption{Confusion matrix from our \textit{contemporaneous} model of news and dislocations, and further statistics.}
\label{confusion-matrix-contemp}
\begin{center}
\begin{tabular}{l|l|c|c|c}
\multicolumn{2}{c}{}&\multicolumn{2}{c}{Predicted $R(t)$}&\\
\cline{3-4}
\multicolumn{2}{c|}{}&0&1&\multicolumn{1}{c}{Total}\\
\cline{2-4}
\multirow{2}{*}{True $R(t)$}& 0 & $148$ & $4$ & $152$\\
\cline{2-4}
& 1 & $20$ & $3$ & $23$\\
\cline{2-4}
\multicolumn{1}{c}{} & \multicolumn{1}{c}{Total} & \multicolumn{1}{c}{$168$} & \multicolumn{    1}{c}{$7$} & \multicolumn{1}{c}{$175$}\\
\end{tabular}
\hspace{1.5cm}
\begin{tabular}{lrrr}
\hline
          &    Precision & Recall &   F1-score   \\
\hline
0  &   0.88 &   0.97 &  0.93  \\
1   &   0.43 &   0.13 &  0.20  \\
Accuracy & &  & 0.86  \\
\hline
\end{tabular}
\end{center}
\end{table}


\begin{table}[h]
\caption{Model for the \textit{prediction} of market dislocations, from the current state of the market and features of news.}
\label{table-model-latex-pred}
\begin{center}
\begin{tabular}{llll}
\hline
Model:              & Logit            & Pseudo R-squared: & 0.342       \\
Dependent Variable: & y                & AIC:              & 283.3737    \\
Date:               & 2023-11-04 13:43 & BIC:              & 294.5248    \\
No. Observations:   & 304              & Log-Likelihood:   & -138.69     \\
Df Model:           & 2                & LL-Null:          & -210.72     \\
Df Residuals:       & 301              & LLR p-value:      & 5.2217e-32  \\
Converged:          & 1.0000           & Scale:            & 1.0000      \\
No. Iterations:     & 17.0000          &                   &             \\
\hline
\end{tabular}
\end{center}

\begin{center}
\begin{tabular}{lrrrrrr}
\hline
            &   Coef. & Std.Err. &       z & P$> |$z$|$ &  [0.025 &  0.975]  \\
\hline
z-vols      &  1.3160 &   0.1595 &  8.2490 &      0.0000 &  1.0033 &  1.6287  \\
eigRatio    & -0.3887 &   0.1610 & -2.4144 &      0.0158 & -0.7043 & -0.0732  \\
n2v-entropy & -0.4155 &   0.1593 & -2.6077 &      0.0091 & -0.7277 & -0.1032  \\
\hline
\end{tabular}
\end{center}
\end{table}

\begin{table}[h!]
\caption{Confusion matrix from our \textit{predictive} model of market dislocations, and further statistics.}
\label{confusion-matrix-pred}
\begin{center}
\begin{tabular}{l|l|c|c|c}
\multicolumn{2}{c}{}&\multicolumn{2}{c}{Predicted $R(t)$}&\\
\cline{3-4}
\multicolumn{2}{c|}{}&0&1&\multicolumn{1}{c}{Total}\\
\cline{2-4}
\multirow{2}{*}{True $R(t)$}& 0 & $150$ & $2$ & $152$\\
\cline{2-4}
& 1 & $13$ & $10$ & $23$\\
\cline{2-4}
\multicolumn{1}{c}{} & \multicolumn{1}{c}{Total} & \multicolumn{1}{c}{$163$} & \multicolumn{    1}{c}{$12$} & \multicolumn{1}{c}{$175$}\\
\end{tabular}
\hspace{1.5cm}
\begin{tabular}{lrrr}
\hline
          &    Precision & Recall &   F1-score   \\
\hline
0  &   0.92 &   0.99 &  0.95  \\
1   &   0.83 &   0.43 &  0.57  \\
Accuracy & &  & 0.91  \\
\hline
\end{tabular}
\end{center}
\end{table}



\section{Conclusions}
\label{sec:conclusionsN}

Starting from a curated selection of economic articles sourced from The Wall Street Journal, our research introduces an innovative and dynamic approach to dissecting news content. We leverage on GPT3.5 to sift out the most salient entities within each article, which become the building blocks of a proposed series of graphs. The graphs track indeed the co-occurrence of such entities among news on a weekly basis, and allow investigations on the inter-relations of topics discussed over time.
Network analysis techniques and fuzzy community detection are then used to design a comprehensive framework, which systematically unveils interpretable topics and surrounding narratives within news.

The importance of the proposed investigations is highlighted by the results of the logistic regression models. Indeed, we test whether there is a statistically significant connection between the features and structure of news, and moments of dislocation within financial markets.
\textcolor{black}{As expected (and desired), lower sentiment within news is more likely to be associated with weeks of market dislocation. However, multiple features computed from our graph construct are found to be also significant, especially from the entropy of discussions and consequent likelihood of contagion of sentiment, both in the contemporaneous and predictive scenarios. This suggests that the interconnectedness of news' topics and structure therein are meaningful aspect to further analyse within financial research, for which our proposed study desires to serve as a first baseline. Improving entity recognition, extending the corpus of news, and designing generalisation studies are examples of possible advances to pursue in this research branch.}

As a final remark, we desire to point to the problem of \textit{network alignment}, which is especially important in network biology \cite{10.3389/fgene.2019.00381}. Many related studies try to find a measure of protein similarity between proteins in different species, since similar protein structures often imply the same biological results. With a parallel approach, one could investigate more deeply whether equivalent structures among news (but that do not account for the actual \say{label} of the topic) result indeed in similar market reactions.




\section*{Acknowledgements}
Deborah Miori's research was supported by the \textit{EPSRC CDT in Mathematics of Random Systems} (EPSRC Grant  EP/S023925/1).

\bibliographystyle{unsrt}  
\bibliography{references}  

\end{document}